\RequirePackage{fix-cm}
\documentclass{svjour3}                     
\smartqed  
\usepackage{graphicx}
\usepackage{bm}
\usepackage{txfonts}
\begin{document}

\title{Theory of Supercurrent Generation in BCS Superconductors
}
\subtitle{ }


\author{Hiroyasu Koizumi 
}


\institute{H. Koizumi \at
              Division of Quantum Condensed Matter Physics, Center for Computational Sciences, University of Tsukuba,Tsukuba, Ibaraki 305-8577, Japan}           

\date{Received: date / Accepted: date}

\maketitle

\begin{abstract}
We revisit the supercurrent generation mechanism for the type of superconductors whose superconducting transition temperature is explained by the BCS theory
(we call it the {\em BCS superconductor}). This revisit is motivated by the reexamination of the ac Josephson effect [H. Koizumi, M. Tachiki, J. Supercond. Nov. Magn. (2015) 28:61] that indicates the charge on the charge carrier for the ac Josephson effect is $q=-e$ (means the electromagnetic vector potential ${\bf A}^{\rm em}$ couples to each electron in the pairing electrons, separately, as $e{\bf A}^{\rm em}$), which strongly suggests that the supercurrent generation mechanism is lacking in the BCS theory since the charge carrier in the BCS theory is the Cooper pair with $q=-2e$ (means ${\bf A}^{\rm em}$ couples to pairing electrons, together, as $2e{\bf A}^{\rm em}$). 

We put forward a possible new supercurrent generation mechanism in the BCS superconductor; we argue that the origin of the supercurrent generation is the emergence of Dirac strings with $\pi$ flux (in the units of $\hbar=1, e=1,c=1$) inside (we call them 
{\em $\pi$-flux Dirac strings}), where the Dirac string is a nodal singularities of the wave function. It appears if the Rashba spin-orbit interaction is added to the BCS model due to its stabilization of the spin-twisting itinerant motion of electrons; then, the $\pi$-flux Dirac string is created as a string of spin-twisting centers. The $\pi$-flux Dirac string generates the cyclotron motion without external magnetic field, and produces topologically protected loop current. A macroscopic persistent current is generated as a collection of such loop currents. 

The above current generation can be also attributed to the emergence of the $U(1)$ instanton of the Berry connection given by ${\bf A}^{\rm fic}=-{ \hbar \over {2e}} \nabla \chi$, ${\bf \varphi}^{\rm fic}={ \hbar \over {2e}} \partial_t \chi$, where $\chi$ is an angular variable of period $2\pi$. 
In other words, the supercurrent is a collective motion produced by the instanton that cannot be reduced to the single particle motion. Then, the appearance of the flux quantum $\Phi_0=h/2e$ and the voltage quantum $V_0=hf/2e$ in the ac Josephson effect ($f$ is the frequency of the radiation field) are explained as topological effects of this instanton. The phase of the macroscopic wave function for the Ginzburg-Landau theory or
the phase of the pair potential of the Bogoliubov-de Gennes equations is identified as $\chi$.

Since the Rashba interaction is absent in the BCS theory, it may be regarded as a weak Rashba interaction limit of the present theory as far as the origin of the phase variable of the macroscopic superconducting wave function is concerned. If the phase variable is treated as a phenomenological parameter, the origin of it does not matter; then, the Ginzburg-Landau theory or the Bogoliubov-de Gennes equations can be used without modification.
 However, the new origin requires the Rashba interaction; thus, the internal electric field for the Rashba interaction is necessary for the occurrence of superconductivity. This may explain the fact that ideal metals like sodium does not show superconductivity since the screening of the electric field is efficient in such materials, suppressing the internal electric field too weak to occur superconductivity. 

\end{abstract}

\section{Introduction}

In the present work, we call the type of superconductors whose superconducting transition temperature is explained by the BCS theory the ``{\em BCS superconductor}''  \cite{BCS1957}. In the BCS superconductor, the superconducting transition temperature is determined by an energy gap formation temperature, where the energy gap is created by the electron pairing due to an effective attractive interaction between electrons that arises from the virtual exchange of phonons. Through the success of the BCS theory, it is now widely-believed that the electron pair formation is the origin of superconductivity. 

As to the practical calculation for phenomena involving supercurrents, the Ginzburg-Landau theory \cite{GL} and the Bogoliubov-de Gennes equations \cite{deGennes} are usually used. In these theoreis, the supercurrent generation is due to the appearance of an angular variable $\phi$ with period $2\pi$ that makes the followings gauge invariant,
\begin{eqnarray}
{\bf A}^{\rm em}-{\hbar \over {2e}}\nabla \phi ; \quad 
\varphi^{\rm em}+{\hbar \over {2e}}\partial_t \phi
\label{potential}
\end{eqnarray}
where $(\varphi^{\rm em}, {\bf A}^{\rm em})$ is the electromagnetic gauge potential ($\varphi^{\rm em}$ and ${\bf A}^{\rm em}$
are scalar and vector potentials, respectively), and the gauge invariance means that the above sums are not affected by the choice of the gauge in $\varphi^{\rm em}$ and ${\bf A}^{\rm em}$ due to compensational changes in $\phi$ \cite{Anderson64,Weinberg}. 
This mode (Nambu-Goldstone mode) was found by Nambu in an effort to rectify the gauge invariance problem of the original BCS paper \cite{Nambu1960} using the generalized Ward-Takahashi identity \cite{Ward,Takahashi57}.
In the BCS superconductors, the required phase $\phi$ appears when the electron pairing is established. It is believe to describe a collective mode of charge $q=-2e$ ($e$ is the absolute value of the electron charge) particle flow \cite{Anderson64,Weinberg,Nambu1960,Bogoliubov58}; $2e$ in the flux quantum $\Phi_0=h/2e$ ($h$ is Planck's constant) and the voltage quantum $V_0=hf/2e$ across the Josephson junction in the presence of a radiation field with frequency $f$, are regarded as due to the pairing electron charge.
It is also considered that $\phi$ is a variable conjugate to the Cooper pair number density $\rho/2$ \cite{Anderson66} ($\rho$ is the electron number density).

Although the origin of the superconductivity due to the electron pairing is believed to be established, the origin of $\phi$ is not. There are more than one theories for the origin of it. The most popular one is the gauge symmetry breaking origin (see for example, Table I and text around it in Ref.~\cite{Anderson}); another competing one is the phase of the Bose-Einstein condensate wave function origin (see for example, Section 2.4 in Ref.~\cite{LeggettBook}). The former uses a particle number non-conserving state as an essential ingredient; however, it suffers from the difficulty in application to fixed particle number systems \cite{LeggettBook,Peierls92} such as isolated superconductors and nuclei in the superconducting states (see for example Ref.~\cite{Ring1980}); note that it is theoretically inconsistent to used the mixed particle number states as the ground state of a fixed particle number system since the Hamiltonian commute with the particle number (i.e., the particle number is a good quantum number and it is fixed).
On the other hand, the latter theory uses a particle number fixed formalism; however, it does not explain the persistent current generation in a natural way, but relies on the topological stability of circular current (or loop current) as an additional requirement \cite{LeggettBook}.

Now, superconductivity of a different type is known in cuprates \cite{Muller1986}. The cuprate superconductors show marked differences from the BCS ones. For example, the superconducting transition temperature is not given by the energy gap formation temperature, but corresponds to the stabilization temperature of coherent-length-sized loop currents for optimally doped samples \cite{Kivelson95}; the coherence-length is in the order of the lattice constant, which is much smaller than that of the BCS superconductor; the normal state from which the superconducting state emerges is not an ordinary metallic state described by the Fermi liquid theory but a doped Mott insulator state; the local magnetic correlation that is a remnant of the parent Mott insulator still exists in the doped compound, giving rise to the hourglass-shaped magnetic excitation spectrum \cite{Neutron}; actually, the magnetic excitations persist entire superconducting hole doping range \cite{MagneticRIXS}, thus, a close relationship between the superconductivity and magnetism is strongly suggested. In spite of all the differences, $\Phi_0=h/2e$ and $V_0=hf/2e$ are observed; thus, it is widely-believed that the origin of the cuprate superconductivity is still the electron pairing. 

The above experiments seem to indicate that the elucidation of the cuprate superconductivity requires a drastic departure from the standard theory. The present author put forward a new theory of superconductivity that dose not contain the pairing-electrons \cite{Koizumi2011,HKoizumi2013,HKoizumi2014,HKoizumi2015} (however, it contains the bipolaron with a hole at each polaron \cite{Muller2007b}; spin-twisting itinerant motion of electrons occurs around each hole). In this theory, the third theory for the origin of $\phi$ is proposed. It uses the Berry phase that was not known during the development of the BCS theory \cite{Berry}. The phase $\phi$ is argued to arise from the singularities of wave functions for spin-twisting itinerant motion of electrons; the centers of spin-twisting creates Dirac strings with $\pi$ flux (in the units of $\hbar=1, e=1,c=1$) inside, and generate $U(1)$ instanton of the Berry connection given by 
 \begin{eqnarray}
 {\bf A}^{\rm fic}=-{ \hbar \over {2e}} \nabla \chi; \quad {\bf \varphi}^{\rm fic}={ \hbar \over {2e}} \partial_t \chi
 \label{fic}
 \end{eqnarray}
where $\chi$ is an angular variable of period $2\pi$ that can be identified as $\phi$. In this theory,  $\chi/2=\phi/2$ is conjugate to the electron number density $\rho$, which differs from the standard theory where $\phi$ is a variable conjugate to the Cooper pair number density $\rho/2$ \cite{Anderson66}.

\begin{figure}
\includegraphics[scale=0.4]{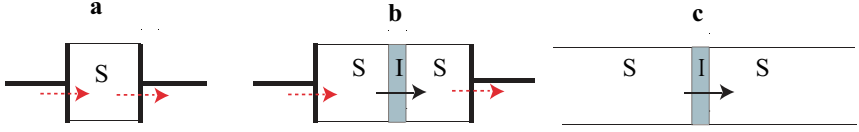}
\caption{ Schematic set-ups for the supercurrent and Josephson effect measurements. $S$ an $I$ indicate superconductor and insulator, respectively. Arrows indicate currents.  {\bf a}: Experimental set-up for supercurrent measurement. {\bf b}: Experimental set-up for Josephson effect measurement.  {\bf c}: Set-up for Josephson effect assumed in the Josephson's derivation.}
\label{Fig1}
\end{figure}

In the presence of ${\bf A}^{\rm fic}$, the effective vector potential for electrons becomes ${\bf A}^{\rm eff}={\bf A}^{\rm em}+{\bf A}^{\rm fic}$, where ${\bf A}^{\rm em}$ is the electromagnetic vector potential.
A macroscopic persistent current is generated as a collection of topologically protected {\em spin-vortex-induced loop currents}.
The appearance of the flux quantum $\Phi_0=h/2e$ and the voltage quantum $V_0=hf/2e$ are explained as topological effects of the $U(1)$ instanton
given in Eq.~(\ref{fic}). 
 One of the advantages of the new theory is that it is formulated in a fixed-particle number formalism, thus, it can be applied to fixed particle number systems without difficulty.
  It also yields a {\em spontaneous feeding current state}, namely, the ground state with energy minima at nonzero values of external current feeding in the situation depicted in Fig.~\ref{Fig1}{\bf a} \cite{Manabe2019}; the value of the spontaneous current depends on the internal state of the superconductor (i.e., the distribution pattern of the spin-vortices and spin-vortex-induce loop currents), thus, the spontaneous feeding current changes flexibly depending on the boundary conditions. This state explains superconductivity, naturally, although such a state has not been obtained by the BCS theory so far. Actually, the inability to obtain such a state is one of the loose ends of the BCS theory \cite{Bloch1966}.

The relevance of the idea presented in the above new theory, {\em supercurrent generation without electron pairing}, to the BCS theory needs to be examined since the origin of the phase variable that produces supercurrent is not settled in the BCS theory, and the Berry phase was not known during the development of the BCS theory. It is noteworthy that it plays a crucial role in explaining the persistent current flow in quantum Hall effects and topological insulators.  Besides, a serious misfit was recently found in the predicted Josephson effect and experimentally observed one \cite{Koizumi2011,HKoizumi2015}, which concerns the boundary conditions for the ac Josephson effect experiment. The boundary condition assumed in the Josephson's predication \cite{Josephson62} and that employed in the real experiment
are actually different (see Figs.~\ref{Fig1}{\bf b} and {\bf c}).
 It is also worth noting that the Josephson's predication assumes a simple appearance of a dc voltage across the Josephson junction, however, a dc voltage does not appear by a simple application of a dc voltage; instead, when a dc voltage is applied, a dc Josephson effect takes over, resulting in a zero voltage across the junction \cite{Shapiro63}. 
 In the experimental situation where a finite voltage exists, there usually exist a radiation field in addition to a dc current feeding from the leads connected to the junction. Since this misfit is the major motivation of the present work, we shall explain it succinctly, below. The details will be revisited in Section \ref{section3.5}.

If we employ the real experimental situation including the current feeding from the leads (the situation in Fig.~\ref{Fig1}{\bf b}), an extra contribution to
 $\dot{\phi}_J$ (denoted by the dotted arrows in Fig.~\ref{Fig1}{\bf b}) arises compared with the Josephson's derivation (Fig.~\ref{Fig1}{\bf c}), where $\phi_J$ is the difference of $\phi$ across the Josephson junction. 
 The two contributions to $\dot{\phi}_J$, one from the chemical potential difference between the leads connected to the junction (the dotted arrows in Fig.~\ref{Fig1}{\bf b}) and the other from the electric field in the non-superconducting region between the two superconductors in the junction (the solid arrow in Fig.~\ref{Fig1}{\bf b}) are equal due to the balance between the voltage from the electric field  in the non-superconducting region and chemical potential difference between those of the two leads connected to the junction.  Thus, the fact that $\dot{\phi}_J={{2eV} \over \hbar}$ is observed experimentally, leads to the conclusion that the carrier charge is $q=-e$ (if we use $q=-2e$ as in the Josephson's prediction, we have $\dot{\phi}_J={{4eV} \over \hbar}$) \cite{Koizumi2011,HKoizumi2015}.
 Although Josephson's predicted relation 
 \begin{eqnarray}
 \dot{\phi}_J={{2eV} \over \hbar}
 \end{eqnarray}
 is valid, it is not due to the electron-pair tunneling in the sense that ${\bf A}^{\rm em}$ couples to pairing electrons, together, as $2e{\bf A}^{\rm em}$. 
Each electron in the pair couples to ${\bf A}^{\rm em}$ as $e{\bf A}^{\rm em}$, and the phase $\phi$ should be attributed to each electron.
In other words,  instead of the standard theory in which $\phi$ is a variable conjugate to the Cooper pair number density $\rho/2$,
we need to adopt the new one where $\phi/2$ is conjugate to the electron number density $\rho$.

Another experiment that suggests the attribution of the phase variable should be to each electron not to each electron-pair comes from the observation of the Josephson effect through the Andreev bound states in the tunneling region with a ring-shaped superconductor under the application of the magnetic field \cite{Spanton:2017aa}. In this experiment, the supercurrent in the
tunneling region are generated by electrons and holes instead of electron pairs, and it is indicated the phase factors $e^{-i \phi/2}$ and $e^{i \phi/2}$ should be attributed to each electron and each hole, respectively, including the contribution from the magnetic flux enclosed by the ring. This separate attribution is in accordance with the new theory in which an effective vector potential ${\bf A}^{\rm eff}={\bf A}^{\rm em}-{ \hbar \over {2e}} \nabla \chi$ is attributed to each charge carrier. 
  
In the present work, we put forward a new supercurrent generation mechanism in the BCS superconductor. It is a similar one developed for the cuprate superconductivity by the present author. In this mechanism, electrons perform spin-twisting itinerant motion stabilized by the Rashba spin-orbit interaction; thus, in order to realize this mechanism, the Rashba spin-orbit interaction needs to be added to the BCS model. Then, cyclotron motion occurs around the singularity of the spin-twisting and loop current produced by it becomes the current element of a macroscopic supercurrent. The line singularities located at the centers of the spin-twisting (they are also centers of the cyclotron motion) form the $\pi$-flux Dirac strings. The appearance of $\Phi_0=h/2e$ and $V_0=hf/2e$ are explained as topological effects of them.
 
Although the new supercurrent generation mechanism presented here is a drastic change from the currently-accepted one, it does not affect the theoretical calculations using the Ginzburg-Landau theory and the Bogoliubov-de Gennes equations if the Rashba interaction is much smaller than the pairing energy gap. In this case, major change is only the re-definition of  the origin of $\phi$. However, the new theory predicts that superconductivity requires the Rashba interaction. This also means that the internal electric field for the itinerant electrons is needed. This may explain the fact that superconductivity does not occur in ideal metals like sodium; in ideal metals, the screening of the internal electric field is efficient, thus, the internal electric field is suppressed; as a consequence, the Rashba interaction is not strong enough to stabilize the spin-twisting itinerant motion.

The organization of the present work is as follows: in Section \ref{section2}, we show that when spin-twisting itinerant motion of electrons is realized the {\em Berry connection for many-body wave functions} is needed in addition to the electron density to obtain the ground state wave function. 
We explain the way to obtain it in the three dimensional system by following the method developed for the two-dimensional case \cite{Manabe2019}.
In Section \ref{section3},
the effective gauge potential in materials, previously introduced, is re-examined for the use in subsequent sections. 
In Section \ref{section3.3} a derivation for the formula for current through Josephson junction is given; here, the number of operator for electrons in the collective mode described by $\chi$ and the number changing operators $e^{\pm {i \over 2}\chi}$ are introduced.
In Section \ref{section3.5}, the ac Josephson effect is revisited by considering the appearance of the Shapiro step \cite{Shapiro63}; the argument starts with the situation where no applied radiation field is present, thus, no voltage across the junction exists; next, a radiation field is applied, and the chemical potential difference appears by the instanton formation. Finally, the establishment of the plateaus in the $I$-$V$ plot (i.e., the Shapiro step) is explained as the consequence of the charging of the junction by treating is as a capacitor.
In Section \ref{section3.7}, the connection between the new theory and standard theory are discussed by employing the number changing operators $e^{\pm {i }\chi}$.
In Section \ref{section5}, the wave packet dynamics of electrons under the influence of the Rashba spin-orbit interaction and magnetic field is studied. We show that the cyclotron motion occurs even without external magnetic field due to the presence of the $\pi$-flux Dirac string. 
In Section \ref{section6}, the gap equation for the new pairing under the influence of the Rashba spin-orbit interaction is considered by assuming that the Rashba interaction is much smaller than the pairing energy gap. We take into account the influence of the Rashba spin-orbit interaction by modifying the pairing states from the original BCS pairing $({\bf k}, \uparrow)$-$(-{\bf k}, \downarrow)$ to $({\bf k}_c, {\bf s}_0({\bf r}_c))$-$(-{\bf k}_c, -{\bf s}_0({\bf r}_c))$ pairing, where ${\bf k}_c$ and ${\bf r}_c$ are the centers of the wave packet in the momentum and coordinate spaces, respectively, and  ${\bf s}_0({\bf r}_c)$ is the direction of spin at ${\bf r}_c$; ${\bf s}_0({\bf r}_c)$ twists along the cyclotron wave packet motion, realizing the spin-twisting cyclotron motion.
In Section \ref{section8}, the modification of the kinetic energy due to the Rashba interaction is derived and the London equation is obtained. It is shown that the state with the spin-twisting cyclotron motion pairing $({\bf k}_c, {\bf s}_0({\bf r}_c))$-$(-{\bf k}_c, -{\bf s}_0({\bf r}_c))$ is more stable that the ordinary pairing $({\bf k}, \uparrow)$-$(-{\bf k}, \downarrow)$.
 In Section \ref{section4}, the problem of the gauge invariance in the BCS theory is revisited.
 Lastly, we conclude the present work in Section \ref{section11}.

\section{Berry Connection for Many-Body Wave Functions and Constraint of the Single-Valued Requirement of the Ground State Wave Function}
\label{section2}

Let us consider the wave function of a system with $N_e$ electrons,
\begin{eqnarray}
\Psi ({\bf x}_1, \cdots, {\bf x}_{N_e},t)
\label{wavef}
\end{eqnarray}
where ${\bf x}_j=({\bf r}_j,s_j) $ denotes the coordinate ${\bf r}_j$  and spin $s_j$ of the $j$th electron.

We define a Berry connection associated with this wave function \cite{Berry}. As will be seen, later, it serves as part of the $U(1)$ gauge field that includes the electromagnetic field for the electrons (see Eqs.~(\ref{effgauge1}) and (\ref{effgauge2})). 

First, we define the parameterized wave function $|n_{\Psi}({\bf r}) \rangle$ with the parameter ${\bf r}$, 
 \begin{eqnarray}
\langle s, {\bf x}_{2}, \cdots, {\bf x}_{N_e} |n_{\Psi}({\bf r},t) \rangle = { {\Psi({\bf r}s, {\bf x}_{2}, \cdots, {\bf x}_{N_e},t)} \over {|C({\bf r} ,t)|^{{1 \over 2}}}}
\end{eqnarray}
where $|C({\bf r} ,t)|$ is the normalization constant given by 
\begin{eqnarray}
|C({\bf r} ,t)|=\int ds d{\bf x}_{2} \cdots d{\bf r}_{N_e}\Psi({\bf r} s, {\bf x}_{2}, \cdots)\Psi^{\ast}({\bf x} s, {\bf x}_{2}, \cdots)
\end{eqnarray}

Using $|n_{\Psi}\rangle$, the {\em Berry Connection for Many-Body Wave Functions} is defined as
 \begin{eqnarray}
{\bf A}^{\rm MB}({\bf r},t)=-i \langle n_{\Psi}({\bf r},t) |\nabla_{\bf r}  |n_{\Psi}({\bf r},t) \rangle
\end{eqnarray}
Here, ${\bf r}$ is regarded as the parameter \cite{Berry}. In the ordinary Hartree-Fock theory, the effect of the Coulomb and exchange interactions from the electron density are taken into account in an average sense; here, we do the same thing for the interaction that affects the phase of the wave function by including the above Berry connection.

We only consider the case where the origin of ${\bf A}^{\rm MB}$ is not the ordinary magnetic field one; thus, we have 
\begin{eqnarray}
\nabla \times {\bf A}^{\rm MB}=0
\label{BMB}
\end{eqnarray}
Then, it can be written in the pure gauge form,
\begin{eqnarray}
 {\bf A}^{\rm MB}=-\nabla \theta
\end{eqnarray}
where $\theta$ is a function which may be multi-valued.

The kinetic energy part of the Hamiltonian is given by
\begin{eqnarray}
K_0={ 1\over {2m}} \sum_{j=1}^{N_e} \left( {\hbar \over i} \nabla_{j} \right)^2
\label{a2}
\end{eqnarray}
where $m$ is the electron mass and $\nabla_{j} $ is the gradient operator with respect to the $j$th electron coordinate ${\bf r}_j$.

Using $\Psi$ and ${\bf A}^{\rm MB}$, we can construct a currentless wave function $\Psi_0$ for the current operator associated with $K_0$
\begin{eqnarray}
\Psi_0 ({\bf x}_1, \cdots, {\bf x}_{N_e},t)=\Psi ({\bf x}_1, \cdots, {\bf x}_{N_e},t)\exp\left(- i \sum_{j=1}^{N_e} \int_{0}^{{\bf r}_j} {\bf A}^{\rm MB}({\bf r}',t) \cdot d{\bf r}' \right)
\label{wavef0}
\end{eqnarray}

In other words, $\Psi ({\bf x}_1, \cdots, {\bf x}_{N_e},t)$ is expressed as
 \begin{eqnarray}
\Psi ({\bf x}_1, \cdots, {\bf x}_{N_e},t)=\Psi_0({\bf x}_1, \cdots, {\bf x}_{N_e},t)\exp\left(- i \sum_{j=1}^{N_e} \theta ({\bf r}_j, t) \right)
\label{f}
\end{eqnarray}
using the currentless wave function $\Psi_0$.

Now consider the situation where the electromagnetic field ${\bf B}^{\rm em}=\nabla \times {\bf A}^{\rm em}$ (${\bf A}^{\rm em}$ is the vector potential) is present.
In this case, the kinetic energy operator is given by
\begin{eqnarray} 
K[{\bf A}^{\rm em}]={ 1\over {2m}} \sum_{j=1}^{N_e} \left( {\hbar \over i} \nabla_{j}-q{\bf A}^{\rm em}({{\bf r}_j}) \right)^2
\label{a3}
\end{eqnarray}
where $q=-e$ is the charge of electron. 

For a while, we consider the case where ${\bf A}^{\rm em} \rightarrow 0$.
The kinetic energy is a functional of ${\bf A}^{\rm em}$ given by
\begin{eqnarray}
E_{\rm kin}=\langle \Psi | K[{\bf A}^{\rm em}] | \Psi \rangle =\langle \Psi_0 | K \left[{\bf A}^{\rm em}+{\hbar \over {q}}  \nabla \theta \right] | \Psi_0 \rangle
\label{Kenergy}
\end{eqnarray}
In the right-most equation, the phase factor $\exp\left(- i \sum_{j=1}^{N_e} \theta ({\bf r}_j, t) \right)$ in $\Psi$ is transferred to the Hamiltonian,
retaining only $\Psi_0$ as the wave function.

 The total energy is a functional of ${\bf A}^{\rm em}$ and $\varphi^{\rm em}$ given by
\begin{eqnarray}
E_{\rm tot}=\langle \Psi | H[{\bf A}^{\rm em}, \varphi^{\rm em}] | \Psi \rangle =\langle \Psi_0 | H \left[{\bf A}^{\rm em}+{\hbar \over {q}}  \nabla \theta ,  \varphi^{\rm em} \right] | \Psi_0 \rangle
\label{Tenergy}
\end{eqnarray}

Now, we treat $\nabla \theta$ as a parameter to be optimized. Let us optimize it by minimizing the total energy $E_{\rm tot}$. This yields 
\begin{eqnarray}
0={{\delta E_{\rm tot}} \over {\delta \nabla \theta}} ={\hbar \over {q}} \left. {{\delta E_{\rm tot}} \over {\delta {\bf A}^{\rm em}} } \right|_{{\bf A}^{\rm em}=0}=-{\hbar \over {q}} {\bf j}
 \label{current0}
\end{eqnarray}
where the relation 
\begin{eqnarray}
  {\bf j}=- {{\delta E_{\rm tot}} \over {\delta {\bf A}^{\rm em}} }
  \label{current}
  \end{eqnarray}
  between the current density ${\bf j}$ and the functional derivative of the total energy with respect to the vector potential is used.
  
 The equation (\ref{current0}) indicates that the energy minimized state is currentless. Thus, if the optimized one is the exact one, it is actually $\Psi_0$ if the ground state is not degenerate. 
  We assume this is the case in the present work. Then, $\Psi_0$ is obtained by the energy minimization.
The fact that ``the energy minimizing ground state is currentless''  is sometimes called the Bloch theorem \cite{Bohm-Bloch}.  $\Psi_0$ satisfies this theorem.  The theory of superconductivity needs to upset this theorem \cite{Bloch1966} to have the current-carrying ground state.

The Bloch theorem can be upset if $\Psi_0$ is multi-valued since in this situation, $\Psi_0$ is not the legitimate wave function (the wave function has to be the single-valued function of the coordinates \cite{Schrodinger}).
If $\Psi_0$  is a real function, only possible multi-valuedness is the sign-change. 
 We call a line of singularities that cause the sign change of the wave function the ``$\pi$-flux Dirac string'', because a Dirac string is a line of singularities of the wave function considered by Dirac \cite{Monopole} and the $\pi$ flux through it (in the units of $\hbar=1, e=1, c=1$) causes the sign change due to the Aharonov-Bohm effect \cite{AB1959}. 

Now we consider the reconstruction of $\Psi$ using $\Psi_0$ that is obtained from the energy minimizing calculation.
First, we note that $\exp(- i \theta )$ must change sign around the $\pi$-flux Dirac string to have the single-valued function $\Psi$. 
This condition can be rephrased using $\chi$ related to $\theta$, 
\begin{eqnarray}\
 \theta={ 1 \over 2}\chi
 \label{chi}
\end{eqnarray}
that the winding number of $\chi$ along path $C$ around $\pi$-flux Dirac string
\begin{eqnarray}
w_C[\chi]={ 1 \over {2 \pi}} \oint_C \nabla \chi \cdot d {\bf r}
\label{w1}
\end{eqnarray}
is an odd integer.

On the other hand, if  $C$ does not encircle the $\pi$-flux Dirac string, we should have
\begin{eqnarray}
w_C[\chi]={ 1 \over {2 \pi}} \oint_C \nabla \chi \cdot d {\bf r}=0
\label{w2}
\end{eqnarray}

We consider the case where the  ``$\pi$-flux Dirac string'', is created by spin-twisting itinerant motion of electrons.
The twisting spin state is expressed using the two-component spin-function
\begin{eqnarray}
 e^{-{ i \over 2} \tau} 
 \left(
 \begin{array}{c}
 e^{i { 1 \over 2} \xi ({\bf r})} \sin {{\zeta ({\bf r}) } \over 2}
 \\
  e^{-i { 1 \over 2} \xi {\bf r})} \cos {{\zeta ({\bf r}) } \over 2} 
 \end{array}
 \right)
 \label{spin-direction}
\end{eqnarray}
where $\zeta$ and $\xi$ are the polar and azimuthal angles of the spin-direction, respectively, and $\tau$ is an angular variable that is introduced to
make the spin-function single valued.

If the Berry connection arise only from this spin-function, we have
\begin{eqnarray}
{\bf A}^{\rm MB}_1=-{1 \over 2} \nabla \tau -{ 1 \over 2} \nabla \xi \cos \zeta
\end{eqnarray}

However, the spin function is that for the opposite spin to the one given in Eq.~(\ref{spin-direction}),
\begin{eqnarray}
 e^{-{ i \over 2} \tau} 
 \left(
 \begin{array}{c}
i e^{i { 1 \over 2} \xi ({\bf r})} \cos {{\zeta ({\bf r}) } \over 2}
 \\
 -i e^{-i { 1 \over 2} \xi {\bf r})} \sin {{\zeta ({\bf r}) } \over 2} 
 \end{array}
 \right)
  \label{spin-direction2}
\end{eqnarray}
we have 
\begin{eqnarray}
{\bf A}^{\rm MB}_2=-{1 \over 2} \nabla \tau +{ 1 \over 2} \nabla \xi \cos \zeta
\end{eqnarray}

Thus, if both spin states with ${\bf A}^{\rm MB}_1$ and ${\bf A}^{\rm MB}_2$ are occupied, 
the over all  Berry connection becomes
\begin{eqnarray}
{\bf A}^{\rm MB}=-{1 \over 2} \nabla \tau
\end{eqnarray}

Actually, the above Berry connection is also obtained in the case with $\zeta=\pi/2$ only from either ${\bf A}^{\rm MB}_1$ or ${\bf A}^{\rm MB}_2$, and we have considered this situation in the cuprate superconductivity \cite{Koizumi2011,HKoizumi2013,HKoizumi2014,HKoizumi2015}.
In any case,  if we have ${\bf A}^{\rm MB}=-{1 \over 2} \nabla \tau$, we can identify $\tau$ as $\chi$, and we consider this case below.

Now the ground state wave function is equipped with the phase $\tau$. Then, we need to have $\tau$ to specify the ground state.
The necessity to have $\tau$ to construct the ground state wave function can be viewed as an extension of the Hohenberg-Kohn theorem  ``the ground state energy is determined by the electron density alone \cite{Hohenberg1964}''. This theorem does not take into account the presence of Dirac strings. If they exist, we need to know $\tau$ in addition. 

We construct $\chi$ using the information on the winding number in Eqs.~(\ref{w1}) and (\ref{w2}), and conservation of local charge as will be explained below.
We first  discretize the three-dimensional continuous space as a cubic lattice of lattice constant $a$ (the volume of the unit cube is $a^3$), which is in the order of the lattice constant of the material.
The electron density $\rho_j$'s and spin-density ${\bf S}_j$'s at the cubic lattice points can be calculated with $\Psi_0$ using only one of the spin functions assuming the electron pair formation with opposite spin states. Here, we need to anticipate the spin-twisting that occurs in the ground state due to the Rashba spin-orbit interaction in obtaining $\Psi_0$. As will be shown later in Section \ref{section8}, such a ground state is really possible. 
However, $\Psi_0$ is a currentless state, thus, the energy gain from the Rashba interaction is absent in $\Psi_0$ even though it exists in $\Psi$.
To find an optimal spin-twisting is a non-trivial problem which we don't know how to do it at present. We simply assume that we have an optimal spin-twisting in the following.

 The system we consider occupies a region of
$N_s$ sites (cubic lattice points) that are composed of $N_c$ cubes (the volume is $N_c a^3$). Each unit cube has 8 sites (or vertices),  6 faces (or plaquettes) , and 12 bonds (or edges), and some of them are shared by other cubes surrounding it.
To obtain $\chi$ means to obtain  $\nabla \chi$ along all bonds. We denote the total number of bonds by $N_b$. 
The value of $\nabla \chi$ along the bond ${k \leftarrow j}$ is written as
\begin{eqnarray}
\tau_{k \leftarrow j}=\chi_k -\chi_j
\end{eqnarray}
To obtain $\chi$, we need to know all $N_b$ values of $\tau_{k \leftarrow j}$'s. Taking $C$ as circumference of each face of the cube, the conditions in Eqs.~(\ref{w1}) and (\ref{w2}) provide $N_f$ equations where $N_f$ is the number fo faces of the cubes in the lattice.

Next, we consider the conditions arising from the conservation of the local charge. According to Eq.~(\ref{current0}), the current through the bond ${k \leftarrow j}$
is given by
\begin{eqnarray}
J_{k \leftarrow j}={{2e} \over \hbar} {{\partial E_{\rm tot}} \over {\partial \tau_{ j \leftarrow i} }}
\end{eqnarray}
Thus, the conservation of charge at site $j$ is given by
\begin{eqnarray}
0= \sum_i {{2e} \over \hbar}  {{\partial E_{\rm tot}} \over {\partial \tau_{ j \leftarrow i}}}+J^{\rm EX}_{j}
\label{Feq3}
\end{eqnarray}
where 
$J^{\rm EX}_{j}$ is the current that is fed externally from the $j$th site.
From Eq.~(\ref{Feq3}), we have $(N_s-1)$ equations, where $N_s$ is the number of sites in the lattice. The subtraction ``$1$'' comes from the fact that the total charge is conserved in the current formalism, thus, the requirement of the conservation at all sites makes one condition redundant.

We impose the condition that when a $\pi$-flux Dirac sting enters a unit cube, it enters through one of the faces of the cube and exits from another one. Then, we have the following equation
\begin{eqnarray}
\nabla \cdot {\bf A}^{\rm MB}=0
\label{condCube}
\end{eqnarray}
for each cube. This condition makes one of the face conditions is redundant for each cube; thus, the conditions from Eqs.~(\ref{w1}) and (\ref{w2}) becomes $(N_f-N_c)$.

The total number of unknowns  is that for $\tau_{k \leftarrow j}$'s of $N_b$ bonds. The equality between the unknowns and the known conditions is given by
\begin{eqnarray}
N_b=(N_s-1)+(N_f-N_c)
\label{Euler}
\end{eqnarray}
Actually, this relation coincides with the Euler's theorem for a three dimensional object.

In this section, we have assumed that the whole system participates the collective motion described by $\chi$. However, this is not correct in general.
We will consider the situation where some electrons perform individual motions in addition to the collective motion described by $\chi$ in Section \ref{section3.7}.

\section{Effective Gauge Potential in Materials}
\label{section3}

Let us derive the equations of motion for $\chi$ and $\rho$. 
We assume that the angular variable $\chi$ is related to the Berry connection as ${\bf A}^{\rm MB}=-{1 \over 2} \nabla \chi$ without assuming 
the presence of Cooper pairs.

To obtain the conjugate momentum of $\chi$, we use the time-dependent variational principle using the following Lagrangian \cite{Koonin1976},
\begin{eqnarray}
{\cal L}\!=\langle \Psi | i\hbar \partial_t \!-\!H[{\bf A}^{\rm em}, \!\varphi^{\rm em}] | \Psi \rangle\!=\!\int \!d{\bf r} \ {{\rho \dot{\chi} \hbar} \over 2} \!+\!i\hbar \langle \Psi_0 |\partial_t| \Psi_0 \rangle
\!-\!E_{\rm tot}\left[{\bf A}^{\rm em}\!+\!{ \hbar \over {2q}}\nabla \chi,\! \varphi^{\rm em}\right]
\label{L}
\end{eqnarray}
where $E_{\rm tot}\left[{\bf A}^{\rm em}\!+\!{ \hbar \over {2q}}\nabla \chi,\! \varphi^{\rm em}\right]
$ is given in Eq.~(\ref{Tenergy}). In this section, we assume the situation where only $\chi$ and its conjugate variable are important dynamical variables.

From the above Lagrangian, the conjugate momentum of $\chi$ is obtained as
\begin{eqnarray}
p_{\chi}= {{\delta {\cal L}} \over {\delta \dot{\chi}}}={\hbar \over {2}}\rho
\label{momentumchi}
\end{eqnarray}
thus, $\chi$ and $\rho$ are canonical conjugate variables apart from some constant.

If we follow the canonical quantization procedure $[\hat{p}_{\chi}({\bf r}, t), \hat{\chi}({\bf r}', t)]=-i\hbar \delta ({\bf r}- {\bf r}')$, where $\hat{p}_{\chi}$ and $\hat{\chi}$ are operators corresponding to ${p}_{\chi}$ and ${\chi}$ respectively, 
we have
\begin{eqnarray}
\left[{ {\hat{\rho}({\bf r}, t)} \over 2} , \hat{\chi}({\bf r}', t) \right]=-i \delta ({\bf r}- {\bf r}')
\end{eqnarray}
where $\hat{\rho}$ is the operator corresponding to $\rho$.

In the standard theory, ${ {\rho({\bf r}, t)} \over 2}$ is attributed to the Cooper pair number density, and $\chi$ is regarded as the canonical conjugate variable to it.  However, we consider it as just a relation between
a collective coordinate $\chi$ and its conjugate variable $\rho$.

Actually, we will re-express it as
\begin{eqnarray}
\left[{ {\hat{\rho}({\bf r}, t)}} , {{\hat{\chi}({\bf r}', t) } \over 2} \right]=-i \delta ({\bf r}- {\bf r}')
\label{cano2}
\end{eqnarray}
and attribute the occurrence of superconductivity as due to the appearance of $\chi /2 $ conjugate to $\rho$.
As shown in Section \ref{section3.5}, this interpretation is more in accordance with the ac Josephson effect.

For simplicity, we only consider the case where  $\langle \Psi_0 |\partial_t| \Psi_0 \rangle=0$ (this will occur if $| \Psi_0 \rangle$ is time-independent or real) is satisfied, below. 

By separating the Coulomb term that is proportional to $\varphi^{\rm em}$, we define $\bar{H}$ as
\begin{eqnarray}
\bar{H}\left[{\bf A}^{\rm em}+{ \hbar \over {2q}}\nabla \chi \right]={H}\left[{\bf A}^{\rm em}+{ \hbar \over {2q}}\nabla \chi, \varphi^{\rm em}\right] -q \int d{\bf r} \ \rho \varphi^{\rm em}
\end{eqnarray}

Then, we define $\bar{E}_{\rm tot}$ by
\begin{eqnarray}
\bar{E}_{\rm tot}\left[{\bf A}^{\rm em}+{ \hbar \over {2q}}\nabla \chi \right]={E}_{\rm tot} \left[{\bf A}^{\rm em}+{ \hbar \over {2q}}\nabla \chi , \varphi^{\rm em}\right]-q \int d{\bf r} \ \rho \varphi^{\rm em}
\end{eqnarray}

Using $\bar{E}_{\rm tot} \left[{\bf A}^{\rm em}+{ \hbar \over {2q}}\nabla \chi \right]$, ${\cal L}$ is written as
\begin{eqnarray}
{\cal L}=-\bar{E}_{\rm tot} \left[{\bf A}^{\rm em}+{ \hbar \over {2q}}\nabla \chi \right]-q \int d{\bf r} \ \rho \left( \varphi^{\rm em}- {\hbar \over {2q}} \dot{\chi} \right) 
\label{Lag}
\end{eqnarray}

The Lagrangian ${\cal L}$ indicates that ${\bf A}^{\rm em}$ and $\varphi^{\rm em}$ always appear in the combinations,
\begin{eqnarray}
{\bf A}^{\rm eff}={\bf A}^{\rm em}+{ \hbar \over {2q}}\nabla \chi
\label{effgauge1}
\end{eqnarray}
and
\begin{eqnarray}
\varphi^{\rm eff}=\varphi^{\rm em}- {\hbar \over {2q}} \dot{\chi}
\label{effgauge2}
\end{eqnarray}
Thus, we may regard $(\varphi^{\rm eff}, {\bf A}^{\rm eff})$ as the basic field instead of $(\varphi^{\rm em}, {\bf A}^{\rm em})$. We call it the {\em effective gauge potential in materials}.

The Hamilton's equations for $\chi$ and $\rho$ are obtained as
\begin{eqnarray}
\dot{\chi}&=&{ {2} \over \hbar} {{\delta {E_{\rm tot}}} \over {\delta \rho}}={ {2} \over \hbar} \left[ {{\delta \bar{E}_{\rm tot}} \over {\delta \rho}} + q\varphi^{\rm em} \right]
\label{H1}
\\
\dot{\rho}&=&{ {2} \over \hbar} \nabla \cdot  {{\delta E_{\rm tot}} \over {\delta \nabla \chi}}={ {2} \over \hbar} \nabla \cdot  {{\delta \bar{E}_{\rm tot}} \over {\delta \nabla \chi}}
\label{H2}
\end{eqnarray}

The equation (\ref{H2}) describes the conservation of the charge 
\begin{eqnarray}
q\dot{\rho}+\nabla \cdot {\bf j}=0
\label{H2'}
\end{eqnarray}
with the current density given by 
\begin{eqnarray}
{\bf j}=-{ {2q} \over \hbar} {{\delta \bar{E}_{\rm tot}} \over {\delta \nabla \chi}}=-{{\delta E_{\rm tot}} \over {\delta {\bf A}^{\rm em}}}
\end{eqnarray}
This indicates that the current density is generated by $\nabla \chi$; in other words, $\chi$ is the collective coordinate that gives rise to supercurrent.

The equation (\ref{H1}) is rewritten as
\begin{eqnarray}
q\varphi^{\rm eff}=- {{\delta \bar{E}_{\rm tot}} \over {\delta \rho}}
\label{cond2}
\end{eqnarray}
This indicates that $-q\varphi^{\rm eff}=e\varphi^{\rm eff}$ plays the role of the chemical potential by taking $\bar{E}_{\rm tot}$ as the total energy.

For a stationary and isolated system, we have $\dot{ \chi}=0$ and $\dot{\rho}=0$.
 From $\dot{ \chi}=0$ and Eq.~(\ref{H1}), we have
 \begin{eqnarray}
 {{\delta E_{\rm tot}} \over {\delta \rho}}=0
 \end{eqnarray}
  This agrees with the condition for the ground state electron density in the density functional theory \cite{Hohenberg1964}. 

Let us consider the gauge invariance problem in $(\varphi^{\rm eff}, {\bf A}^{\rm eff})$. In classical theory, the gauge invariance is the invariance for
the electric field ${\bf E}^{\rm em}$ and the magnetic field ${\bf B}^{\rm em}$ 
\begin{eqnarray}
{\bf E}^{\rm em}= -\partial_t {\bf A}^{\rm em} - \nabla \varphi^{\rm em}; \quad {\bf B}^{\rm em}=\nabla \times {\bf A}^{\rm em}
\end{eqnarray}
with respect to the following modifications,
\begin{eqnarray}
{\bf A}^{\rm em} \rightarrow {\bf A}^{\rm em}-{\hbar \over {2q}}\nabla \phi ; \quad \varphi^{\rm em} \rightarrow
\varphi^{\rm em}+{\hbar \over {2q}}\partial_t \phi
\label{gauge1}
\end{eqnarray}

In quantum mechanics, the gauge transformation requires an additional change in the phase of the wave function for the material interacting with the electromagnetic field 
\begin{eqnarray}
\psi({\bf x}, t) \rightarrow e^{-{i \over 2} \phi} \psi({\bf x}, t) 
\label{gauge2}
\end{eqnarray}
This means that we need to adjust the $U(1)$ phase factor of the wave function in response to the change of the gauge.
If this adjustment is not properly done, a surplus whole system motion appears since the $U(1)$ phase factor also describes a whole system motion.

In the present theory, the gauge invariant ${\bf A}^{\rm eff}$ is obtained from the single-valuedness of the wave function, and
the conservation of the local charge.  
Then, by substituting ${\bf A}^{\rm eff}$ in Eq.~(\ref{cond2}), the gauge invariant $\varphi^{\rm eff}$ is obtained.
Here, the arbitrariness in gauge chosen for $\varphi^{\rm em}$ is absorbed in the arbitrariness of $\partial_t \chi$. Therefore, we can obtain the gauge invariant $(\varphi^{\rm eff}, {\bf A}^{\rm eff})$. This also means that if we stick to $(\varphi^{\rm eff}, {\bf A}^{\rm eff})$, the surplus whole system motion does not appear since the relation between the gauge of the gauge potential and the phase factor on the wave function is intact.

Let us see that the phase change in Eq.~(\ref{gauge2}) in the wave function can be obtained as a particular case for the above mentioned evaluation of $\chi$ that satisfies  the single-valuedness of the wave function, and
the conservation of the local charge. First, we assume $\Psi_0$ in Eq.~(\ref{f}) is the exact solution for the first chosen $(\varphi^{\rm em}, {\bf A}^{\rm em})$. Then, 
the fact that $\Psi_0$ is optimized for the first chosen $(\varphi^{\rm em}, {\bf A}^{\rm em})$ means that, for the gauge transformation in Eq.~(\ref{gauge1}), the solution $\chi$ evaluated by the single-valuedness of the wave function, and
the conservation of the local charge yields $\chi=\phi$ within an arbitrary constant.
This is because the gauge invariant $(\varphi^{\rm eff}, {\bf A}^{\rm eff})$ is obtained as
\begin{eqnarray}
{\bf A}^{\rm eff}= {\bf A}'^{\rm em}+{ \hbar \over {2q}}\nabla \chi={\bf A}^{\rm em}-{\hbar \over {2q}}\nabla \phi+{ \hbar \over {2q}}\nabla \chi
\\
\varphi^{\rm eff}= \varphi'^{\rm em}-{ \hbar \over {2q}}\partial_t \chi=\varphi^{\rm em}+{\hbar \over {2q}}\partial_t\phi-{ \hbar \over {2q}}\partial_t \chi
\label{chosengauge}
\end{eqnarray}
and $\Psi_0$ is optimized for $(\varphi^{\rm em}, {\bf A}^{\rm em})$ means $(\varphi^{\rm eff}, {\bf A}^{\rm eff})=(\varphi^{\rm em}, {\bf A}^{\rm em})$;
thus, we have $\nabla \phi=\nabla \chi$ and $\partial_t \phi=\partial_t \chi$. 

\section{A derivation for the formula for current through Josephson junction}
\label{section3.3}

In this section we derive the formula for the current flow through the Josephson junction including the leads connected to it.

Let us construct boson field operators from Eq.~(\ref{cano2})
\begin{eqnarray}
\hat{\psi}_{e}^{\dagger}({\bf r})= \left(\hat{\rho}({\bf r}) \right)^{1/2} e^{i {\hat{\chi}({\bf r}) \over 2}}, \quad \hat{\psi}_{e}({\bf r})= e^{-i { \hat{\chi}\over 2}}\left( \hat{\rho}({\bf r}) \right)^{1/2}, \quad [\hat{\psi}_{e}({\bf r}),\hat{\psi}_{e}^{\dagger}({\bf r}')]=\delta({\bf r}-{\bf r}')
\label{boson1}
\end{eqnarray}

 Using the above boson field operators, we construct the number operators for electrons participating in the collective mode described by $\chi$ in $S_L$ and $S_R$ ($\hat{N}_L$, $\hat{N}_R$, respectively), creation operators ($\hat{C}^{\dagger}_L$, $\hat{C}^{\dagger}_R$, respectively), and annihilation operators ($\hat{C}_L$, $\hat{C}_R$, respectively), as follows
 \begin{eqnarray}
 \hat{C}^{\dagger}_j=\int_{S_j} d{\bf r} \hat{\psi}_{e}^{\dagger}({\bf r}) 
, \quad \hat{C}_j=\int_{S_j} d{\bf r} \hat{\psi}_{e}^{}({\bf r}),  \quad \hat{N}_j= \hat{C}^{\dagger}_j  \hat{C}_j, \quad j=L,S
\label{Cj}
 \end{eqnarray}
 
 They satisfy the boson commutation relation
 \begin{eqnarray}
 \quad [\hat{C}_j, \hat{C}^{\dagger}_k]=\delta_{jk}
 \end{eqnarray}
 
 Through the creation and annihilation operators, the phase operators $\hat{\chi}_j$ that are conjugate to the number operators $\hat{N}_j$
 are defined as
  \begin{eqnarray}
 \hat{C}^{\dagger}_j=(\hat{N}_j)^{ 1 \over 2} e^{ { i \over 2} \hat{\chi}_j}
, \quad \hat{C}_j=e^{- { i \over 2} \hat{\chi}_j}(\hat{N}_j)^{ 1 \over 2} , \quad j=L,S
\label{chij}
 \end{eqnarray}
 Strictly speaking, $\hat{\chi}$ is not a hermitian operator \cite{Fujikawa2004}; however, we treat it as hermitian by neglecting a minor difference.
 
 The important relation for the later discussion is following
   \begin{eqnarray}
[e^{- { i \over 2} \hat{\chi}_j}, \hat{N}_j]=e^{- { i \over 2} \hat{\chi}_j}
\label{commchi}
 \end{eqnarray}
 
 Let us define eigenstates of $\hat{N}_j$ and $\hat{\chi}_j$ as 
    \begin{eqnarray}
\hat{N}_j | N_j \rangle =N_j | N_j \rangle, \quad  e^{{ i \over 2} \hat{\chi}_j} | {\chi}_j \rangle =e^{ { i \over 2} {\chi}_j} | {\chi}_j  \rangle
\label{commchi2}
 \end{eqnarray}
 
 Then, from Eqs.~(\ref{commchi}) and (\ref{commchi2}), we have
 \begin{eqnarray}
e^{\pm { i \over 2}\hat{\chi}_j }| N_j \rangle =| N_j \pm 1\rangle, \quad e^{\pm{ i \over 2}{\chi}_j } \langle {\chi}_j| N_j \rangle=\langle {\chi}_j| N_j \pm 1 \rangle
\label{N-chi}
 \end{eqnarray}

The standard form of the energy operator for the Josephson junction is given by 
\begin{eqnarray}
H_J= E_{J0} \left( \hat{C}^{\dagger}_L \hat{C}_R +\hat{C}^{\dagger}_R \hat{C}_L \right)
\end{eqnarray}
where $E_{J0}$ is a constant \cite{Nori2017}, but we include the effect of the current feeding from the leads as
\begin{eqnarray}
H'_J= \sum_{\sigma \sigma'}E'_{J0} \left( c_{l \sigma}^{\dagger} c_{L \sigma} \hat{C}^{\dagger}_L \hat{C}_R c_{R \sigma'}^{\dagger} c_{r \sigma'} 
+c_{r \sigma}^{\dagger} c_{R \sigma}\hat{C}^{\dagger}_R \hat{C}_L c_{L \sigma'}^{\dagger} c_{l \sigma'} \right)
\end{eqnarray}
where $c_{l \sigma}^{\dagger}, c_{r \sigma}^{\dagger}, c_{L \sigma}^{\dagger}$, and $c_{R\sigma}^{\dagger}$ ($c_{l \sigma}, c_{r \sigma}, c_{L \sigma}$, and $c_{R\sigma}$) are 
creation (annihilation) operators of electrons with spin $\sigma$ in the left lead, right lead, left superconductor, and right superconductor, respectively.

We denote the junction state as $|\chi_L, \chi_R \rangle$. From Eq.~(\ref{N-chi}), the matrix elements of  $\hat{C}^{\dagger}_L \hat{C}_R$ and 
 $\hat{C}^{\dagger}_R \hat{C}_L$ are shown to be diagonal with diagonal elements
 \begin{eqnarray}
\langle \chi_L, \chi_R |  \hat{C}^{\dagger}_L \hat{C}_R |\chi_L, \chi_R \rangle=e^{- {i \over 2}(\chi_L -\chi_R)}, \ 
\langle \chi_L, \chi_R |  \hat{C}^{\dagger}_R \hat{C}_L |\chi_L, \chi_R \rangle=e^{ {i \over 2}(\chi_L -\chi_R)} 
 \end{eqnarray}
 Actually, physically meaning quantity is the relative phase $(\chi_L -\chi_R)$, thus we may write the junction state as
 $|\chi_L - \chi_R \rangle$. 
 Now we denote the state vector for the (junction $+$ leads) system as $|\chi_L-\chi_R, S_l, S_r \rangle$, where
$ S_l$ and $S_r$ labels for the left-lead state and right-lead state, respectively. 

We also replace ${1\over {2}} \nabla \chi$ by the gauge invariant $-{e\over {\hbar}}{\bf A}^{\rm eff}$, yielding
 \begin{eqnarray}
\langle \chi_L -\chi_R, S_l, S_r|  \hat{C}^{\dagger}_L \hat{C}_R |\chi_L - \chi_R, S_l, S_r \rangle=\exp  \left(i{ e \over \hbar} \int^L_R {\bf A}^{\rm eff} \cdot d{\bf r} \right)
 \end{eqnarray}

Then, the junction energy is calculated as
\begin{eqnarray}
E_J&=& \langle G | H_J' | G \rangle
\nonumber
\\
 &=& E'_{J0}\sum_{\sigma, \sigma', S_l, S_r} \langle G| c_{l \sigma}^{\dagger} c_{L \sigma}|\chi_L - \chi_R, S_l, S_r \rangle
e^{ i{ e \over \hbar} \int^L_R {\bf A}^{\rm eff} \cdot d{\bf r}}
\langle \chi_L -\chi_R, S_l, S_r| c_{L \sigma'}^{\dagger} c_{l \sigma'}|G \rangle
\nonumber
\\
&+& {\rm c.c.}
\nonumber
\\
&=&2E_{JJ} \cos  \left({ e \over \hbar} \int^L_R {\bf A}^{\rm eff} \cdot d{\bf r} + \alpha \right)
\end{eqnarray}
where $\alpha$ is the phase of the following constant
\begin{eqnarray}
C_J= E'_0\sum_{\sigma, \sigma', S_l, S_r} \langle G| c_{l \sigma}^{\dagger} c_{L \sigma}|\chi_L - \chi_R, S_l, S_r \rangle
\langle \chi_L -\chi_R, S_l, S_r| c_{L \sigma'}^{\dagger} c_{l \sigma'}|G \rangle =|C_J|e^{i \alpha}
\end{eqnarray}
and $E_{JJ}=|C_J|$.

From $E_J$, the current through the junction is obtained as
\begin{eqnarray}
J_{\rm ac}= {{2 e E_{JJ}} \over \hbar}  \sin  \left(-{ e \over \hbar} \int_L^R {\bf A}^{\rm eff} \cdot d{\bf r} +\alpha \right)
\label{JJ}
\end{eqnarray}
This is the standard form of the Josephson current when the Josephson junction is used as a circuit element \cite{Nori2017}.

\section{Revisiting ac Josephson effect}
\label{section3.5}
We revisit the ac Josephson effect problem in this section. This is a modified and enlarged version of our previous work \cite{HKoizumi2015}.

Let us denote two superconductors in the Josephson junctions as S$_L$ and S$_R$. The angular variable $\chi$ is assumed to be continuous
along the line connecting S$_L$ and S$_R$ (we take it in the $x$-direction); values of $\chi$ on S$_L$ and S$_R$ are denoted as $\chi_L$ and $\chi_R$, respectively. Then, according to Eq.~(\ref{current}) the current-flow through the junction is a function of
\begin{eqnarray}
\int_L^R {\bf A}^{\rm eff} \cdot d{\bf r}=\int_L^R {\bf A}^{\rm em} \cdot d{\bf r}+{\hbar \over {2q}}(\chi_R-\chi_L)
\end{eqnarray}
This formula may be regarded as a sum of the phase due to the Peierls substitution of the transfer integral between S$_L$ and S$_R$, and the phase from 
the wave functions (Eq.~(\ref{f}) with $\theta={ 1 \over 2}\chi_L$ on S$_L$  and $\theta={ 1 \over 2}\chi_R$ on S$_R$). The important point is that the gauge invariant ${\bf A}^{\rm eff}$ appears instead of ${\bf A}^{\rm em}$.

Since the change of $\chi_R \rightarrow \chi_R+ 4 \pi n$ ($n$ is an integer) or $\chi_L \rightarrow \chi_L+ 4 \pi n$ ($n$ is an integer) does not change the wave functions
on the superconductors, 
the current is a function of the angular variable 
\begin{eqnarray}
{q \over \hbar} \int_L^R {\bf A}^{\rm eff} \cdot d{\bf r} 
\label{phiJ}
\end{eqnarray}
with period $2\pi$ \cite{Weinberg}.
The current through the junction is often approximated as
\begin{eqnarray}
J_{\rm ac}=J_c \sin \phi_J
\label{sin}
\end{eqnarray}
where $\phi_J$ is given by
\begin{eqnarray}
\phi_J={q \over \hbar} \int_L^R {\bf A}^{\rm eff} \cdot d{\bf r}  + \alpha
\end{eqnarray}
as is given in Eq.~(\ref{JJ}), but we do not assume the above form in the following unless otherwise stated.

According to Eq.~(\ref{cond2}), the chemical potential $\mu$ is obtained as 
\begin{eqnarray}
\mu=-q\varphi^{\rm eff}
\end{eqnarray}
It is assumed to be continuous along the junction.
 
From Eq.~(\ref{effgauge2}), the difference of the chemical potential on S$_L$ and on S$_R$ is given by
\begin{eqnarray}
\int_L^R \nabla \mu \cdot d{\bf r}=-q\int_L^R \nabla \varphi^{\rm em}\cdot d{\bf r} +{\hbar \over {2}}\int_L^R \nabla \dot{\chi}\cdot d{\bf r}=\mu_R-\mu_L
\label{chemdif}
\end{eqnarray}
where $\mu_L$ and $\mu_R$ are chemical potentials of S$_L$ and S$_R$, respectively.
 When the radiation field is absent, we have $\mu_L=\mu_R$ and the dc Josephson effect occurs.

Let us apply a radiation field with frequency $f$. 
Then, $\nabla \varphi^{\rm em}$ arises from this radiation field, which oscillates with frequency $f$; thus, its time average over the interval $f^{-1}$ is zero. 
Since the current is dc we have $\partial_t {\bf A}^{\rm eff}=0$ from Eqs.~(\ref{phiJ}) and (\ref{sin}).
Then, using $\partial_t {\bf A}^{\rm eff}=0$ and the fact that ${\bf A}^{\rm em}$ oscillates with frequency $f$, the time average of $\partial_t \nabla \chi$ over the interval $f^{-1}$ is calculated to be zero.

Then, using Eq.~(\ref{chemdif}) and the fact that the time average of $\partial_t \nabla \chi$ over the interval $f^{-1}$ is zero, 
the chemical potential difference averaged over time interval $0 < t  <f^{-1}$ is calculated as 
\begin{eqnarray}
\mu_R-\mu_L
= {{\hbar f} \over {2}}  \int _0^{f^{-1}}dt \int_L^R \partial_x \partial_t{\chi} d x={{\hbar f} \over {2}}  \int _0^{f^{-1}}dt \int_L^R (\partial_x \partial_t-\partial_t \partial_x ){\chi} d x={{hf} \over {2}} n
\label{voltage}
\end{eqnarray}
 where  $\partial_t \partial_x \chi$ is added in going from the left of the second equality to the right since its time-average is zero. 
 
 When a singularity of $\chi$ (``instanton")  is created, nonzero $n$ arises, where $n$ is  
\begin{eqnarray}
 n={ 1 \over {2\pi}} \int _0^{f^{-1}}dt \int_L^R (\partial_x \partial_t-\partial_t \partial_x ){\chi} d x= { 1 \over {2\pi}} \oint _{
 \partial \{ [0,{f^{-1}}] \times [L,R] \} } d \chi
  \end{eqnarray}
  the winding number of $\chi$ along boundary of integration.
   This indicates that the chemical potential difference ${{hf} \over {2}} n$ arises due to the creation of the ``instanton". This instanton may be viewed as a flow of a vortex in the interface region of the two superconductors.
  
 Next we consider the situation where a chemical potential difference appears due to the instanton creation. Due to the fact that the junction is a capacitor, the chemical potential difference is balanced by the electric field ${\bf E}^{\rm em}$ in the insulator region generated by charging of the capacitor. 
 
 Let us calculate $ \phi_J$ for this state. We take the time derivative of $ \phi_J$ in Eq.~(\ref{phiJ}),
 \begin{eqnarray}
\dot{ \phi}_J&=&{q \over \hbar} \int_L^R \dot{\bf A}^{\rm em} \cdot d{\bf r}-{1 \over 2} \int_L^R \nabla \dot{\chi} \cdot d{\bf r}
\nonumber
\\
&=&-{q \over \hbar} \int_L^R {\bf E}^{\rm em} \cdot d{\bf r}-{q \over \hbar}\int_L^R \nabla \varphi^{\rm eff} \cdot d{\bf r}
\nonumber
\\
&=&-{q \over \hbar} \int_L^R {\bf E}^{\rm em} \cdot d{\bf r}+{{\mu_R-\mu_L} \over \hbar}
 \end{eqnarray}
 where the relation ${\bf E}^{\rm em}= - \partial_t {\bf A}^{\rm em}-\nabla \varphi^{\rm em}$ is used.
 
 The balance of the chemical potential difference and the electric field in the insulator region of the junction requires
  \begin{eqnarray}
\mu_R-\mu_L=-q\int_L^R {\bf E}^{\rm em} \cdot d{\bf r}=qV
 \end{eqnarray}
 where $V$ is the voltage across the junction.
 
 Thus, we have
  \begin{eqnarray}
\dot{ \phi}_J={{2q} \over \hbar}V=-{{2e} \over \hbar}V
\label{phidot}
 \end{eqnarray}
 using $q=-e$. This is the Josephson relation. 
 Actually, ${\bf E}^{\rm em}$ contains a contribution from the radiation field with frequency $f$; however, it does not change the average voltage $V$.
 Thus, this relation is valid in this averaged sense.
 
The fact that the Josephson relation is obtained using $q=-e$ means that ${\bf A}^{\rm em}$ couples to each electron in the pairing electrons, separately, as $e{\bf A}^{\rm em}$. This contradicts the standard theory in which ${\bf A}^{\rm em}$ couples to pairing electrons, together, as $2e{\bf A}^{\rm em}$ \cite{Josephson62}. Note that, for the Bogoliubov quasiparticle, $q=-e$ means ${\bf A}^{\rm em}$ couples to the electron and hole parts of it as $e{\bf A}^{\rm em}$ and $-e{\bf A}^{\rm em}$, respectively. This smoothly connects to the coupling observed in the Andreev bound state in the tunneling region of the ring shaped Josephson junction \cite{Spanton:2017aa}.
 
 The presence of the radiation field with frequency $f$ enables the flow of dc current if the resonance condition 
 \begin{eqnarray}
{{2e} \over \hbar}V=2\pi f n
\label{Vcond}
 \end{eqnarray}
 is satisfied, where $n$ is an integer. 
 This relation is equal to the one in Eq.~(\ref{voltage}), and gives rise to the voltage quantization
  \begin{eqnarray}
V={{hf} \over {2e}}n
 \end{eqnarray}
 observed as ``Shapiro steps'' \cite{Shapiro63}.
 
 Let us examine this Shapiro step problem by adopting the approximate current expression in Eq.~(\ref{sin}). 
By setting $V$ in Eq.~(\ref{phidot}) as $V_0+V_1 \cos \omega t, \omega=2\pi f$, we have
\begin{eqnarray}
\dot{\phi}_J={{2qV_0} \over \hbar} + {{2qV_1} \over {\hbar }} \cos \omega t.
\end{eqnarray} 
Then, $\phi_J$ is calculated as
\begin{eqnarray}
\phi_J={{2qV_0} \over \hbar} t +{{2qV_1} \over {\hbar \omega}} \sin \omega t +\gamma
\end{eqnarray}

Substituting the above $\phi_J$ in Eq.~(\ref{sin}), we obtain the following well-known current expression
\begin{eqnarray}
J_{\rm ac}=J_c \sum_{n=-\infty}^{\infty} J_n \left({{2qV_1} \over {\hbar \omega}} \right)
\sin \left(  {{2qV_0} \over \hbar} t +n\omega t +\gamma \right)
\label{J1}
\end{eqnarray}
where $J_n(x)$ is the Bessel function.

The dc current $\bar{J}_{\rm ac}$ flow occurs when the condition 
\begin{eqnarray}
{{2qV_0} \over \hbar}  +n\omega=0
\label{J2}
\end{eqnarray}
is fulfilled \cite{TinkhamText}.  This is equivalent to the condition in Eq.~(\ref{Vcond}).

When an oscillating electric field with frequency $\omega = {{2qV_0} \over {\hbar n}}$ ($n$ is an integer) is applied, the voltage
\begin{eqnarray}
V_n={{\hbar \omega } \over {2e} }n={{hf } \over {2e} }n
\end{eqnarray}
appears.
 
Let us consider the charging of the junction. We denote the capacitance of the junction as $C_J$. Then, the charge $\pm Q$ stored in the junction is given by
\begin{eqnarray}
Q=C_J V_n.
\label{J3}
\end{eqnarray}

We consider the case where the junction is not a perfect capacitor. Then, the tunneling causes the discharging by the recombination of the opposite charges across the insulator. The equation for this process is described by
\begin{eqnarray}
{{dQ} \over {dt}}=-\alpha_d Q,
\end{eqnarray}
 where $\alpha_d$ is the discharging rate.
By including the current flow due to the tunneling $\bar{J}_{\rm ac}$ and the current fed from the lead $I$, the conservation of the charge is given by
\begin{eqnarray}
{{dQ} \over {dt}}=I-\bar{J}_{\rm ac}-\alpha_d Q.
\end{eqnarray}

From the stationary condition ${{dQ} \over {dt}}=0$ and Eqs.~(\ref{J1}), (\ref{J2}), and (\ref{J3}) with $-1 \leq \sin \gamma \leq 1$, we have 
\begin{eqnarray}
\alpha_d C_J V_n -J_{c}J_n \left( {{2e V_1} \over {\hbar \omega}} \right)\le I \le \alpha_d C_J V_n+J_{c}J_n \left( {{2e V_1} \over {\hbar \omega}} \right),
\end{eqnarray}
where $n\ge0$ is assumed.
The above $I-V$ characteristic is the Shapiro step observed in the experiment \cite{Shapiro68}. 

Note that the applied radiation field actually plays two roles; one is the creation of the instanton that generates the chemical potential difference given in Eq.~(\ref{voltage}), and the other is the maintenance of the dc voltage by the resonance condition in Eq.~(\ref{Vcond}).

\section{The number changing operator $e^{-i \hat{\chi}}$ and the BCS theory}
\label{section3.7}

In this section, we explore a connection between the new theory and standard theory.

Let us briefly review the BCS theory \cite{BCS1957}. The model Hamiltonian is given by $H_{\rm kin}+H_{\rm int}$, where $H_{\rm kin}$ is the kinetic energy given by
\begin{eqnarray}
H_{\rm kin}= \sum_{{\bf k} \sigma} \xi_0({\bf k})c^{\dagger}_{{\bf k} \sigma}c_{{\bf k} \sigma}
\end{eqnarray}
$\xi({\bf k})$ is the energy measured from the Fermi energy ${\cal E}_{F}$ given by
\begin{eqnarray}
\xi_0({\bf k})={\cal E}({\bf k})-{\cal E}_{F}
\end{eqnarray}
and $H_{\rm int}$ is the interaction energy given by
\begin{eqnarray}
H_{\rm int}=\sum_{{\bf k} {\bm \ell}} V_{{\bf k} {\bm \ell}}c^{\dagger}_{{\bf k} \uparrow}c^{\dagger}_{-{\bf k} \downarrow}c_{-{\bm \ell} \downarrow}c_{{\bm \ell} \uparrow}.
\label{Hint}
\end{eqnarray}
The electron pairing occurs between electrons near the Fermi surface since attractive $V_{{\bf k} {\bm \ell}}$ only exists in that region.
In the BCS interaction, $V_{{\bf k} {\bm \ell}}$ is nonzero ($V_{{\bf k} {\bm \ell}}=-g$) only when $|{\xi}_0({\bf k})|, |{\xi}_0({\bm \ell})| <  \hbar \omega_D$ ($\omega_D$
 is the Debye frequency) is satisfied. Then, $\Delta_{\bf k}$ becomes independent of ${\bf k}$, and we express it as $\Delta$.

The superconducting state is given by the following state vector,
\begin{eqnarray}
|{\rm BCS} \rangle=\prod_{\bf k}(u_{\bf k}+v_{\bf k}c^{\dagger}_{{\bf k} \uparrow}c^{\dagger}_{-{\bf k} \downarrow})
|{\rm vac} \rangle.
\label{BCS}
\end{eqnarray}
This state exploits the attractive interaction between electron pairs $({\bf k} \uparrow)$ and $(-{\bf k} \downarrow)$ and the following energy gap equation is 
obtained,
\begin{eqnarray}
\Delta=g\sum_{ |{\xi}_0({\bm \ell})| <  \hbar \omega_D } u_{\bm \ell}v_{\bm \ell}
\end{eqnarray}
 and
 $u_{\bf k}$ and $v_{\bf k}$ are parameters given using $\Delta$ and $\xi({\bf k})_0$ as
\begin{eqnarray}
u_{\bf k}={1 \over \sqrt{2}} \left(1 + {{\xi_0({\bf k})} \over \sqrt{\xi_0^2({\bf k})+\Delta^2}} \right)^{1/2}
\end{eqnarray}
and
\begin{eqnarray}
v_{\bf k}={1 \over \sqrt{2}} \left(1 -{{\xi_0({\bf k})} \over \sqrt{\xi_0^2({\bf k})+\Delta^2}} \right)^{1/2},
\end{eqnarray}
respectively.

The total energy by the formation of the energy gap is given by
\begin{eqnarray}
E_{\rm s}^{\rm BCS}&=&E_{\rm n}^{\rm BCS}-{1 \over 2} N(0)\Delta^2
\end{eqnarray}
 where $E_{\rm n}^{\rm BCS}$ is the normal state energy, and $N(0)$ is the density of states at the Fermi energy\cite{BCS1957}.

For the BCS theory, we can obtain the relation in Eq.~(\ref{momentumchi}) as follows;
let us express the BCS state for the coarse-grained cell (its volume is unity) with the center position ${\bf r}$ as
\begin{eqnarray}
|\Psi_{\rm BCS} ({\bf r},t) \rangle=\prod_{\bf k} \left( \sin \theta_{\bf k} ({\bf r}) + e^{-i \chi({\bf r},t) } \cos \theta_{\bf k} ({\bf r}) c^{\dagger}_{{\bf k} \uparrow}c^{\dagger}_{-{\bf k} \downarrow} \right) |{\rm vac} \rangle
\end{eqnarray}

Then, the Lagrangian corresponding to Eq.~(\ref{L}) is given by
\begin{eqnarray}
{\cal L}_{\bf BCS}\!=\int d{\bf r} \langle \Psi_{\rm BCS} ({\bf r},t)  | i\hbar \partial_t \!-\!H_{\rm BCS}| \Psi_{\rm BCS} ({\bf r},t)  \rangle\!=\!\int \!d{\bf r} \ {{\rho \dot{\chi} \hbar} \over 2}\!-\! \int d{\bf r}\langle \Psi_{\rm BCS} ({\bf r},t)  | H_{\rm BCS} | \Psi_{\rm BCS} ({\bf r},t)  \rangle\!
\nonumber
\\
\label{L2}
\end{eqnarray}
where $H_{\rm BCS}$ is $H_{\rm kin}+H_{\rm int}$ in the corse-grained cell centered at ${\bf r}$, and the relation
\begin{eqnarray}
\rho({\bf r})=2\sum_{\bf k} \cos^2 \theta_{\bf k} ({\bf r}) 
\end{eqnarray}
is used. 

From Eq.~(\ref{L2}), we obtain $p_{\chi}=\hbar \rho /2$ as in Eq.~(\ref{momentumchi}). Thus, we may construct the following boson field operators
\begin{eqnarray}
\hat{\psi}_{2e}^{\dagger}({\bf r})= \left({ \hat{\rho}({\bf r}) \over 2} \right)^{1/2} e^{-i { \hat{\chi}({\bf r}) }}, \quad \hat{\psi}_{2e}({\bf r})= e^{i { \hat{\chi} }}\left({ \hat{\rho}({\bf r}) \over 2} \right)^{1/2}, \quad [\hat{\psi}_{2e}({\bf r}),\hat{\psi}_{2e}^{\dagger}({\bf r}')]=\delta({\bf r}-{\bf r}')
\label{boson2}
\end{eqnarray}

It is tempting to associate $\hat{\psi}_{2e}({\bf r})$ to the electron-pair field operator $\hat{\psi}_{\uparrow}({\bf r}) \hat{\psi}_{\downarrow}({\bf r})$ ($\hat{\psi}_{\sigma}({\bf r})$ is the electron field operator with spin $\sigma$) since ${\rho({\bf r}) \over 2}$ can be considered as the electron-pair number density; however, such an association is invalid since the latter field operator dose not satisfy the boson commutation relation,
\begin{eqnarray}
 [\hat{\psi}_{\uparrow}({\bf r}) \hat{\psi}_{\downarrow}({\bf r}),\hat{\psi}^{\dagger}_{\downarrow}({\bf r}') \hat{\psi}^{\dagger}_{\uparrow}({\bf r}')] \neq \delta({\bf r}-{\bf r}')
\end{eqnarray}

Actually, we should use the field operators in Eq.~(\ref{boson1}) instead of Eq.~(\ref{boson2}) since the misfit in the ac Josephson effect indicates that the collective motion for the supercurrent are those give in Eq.~(\ref{boson1}).

As in Eqs.~(\ref{Cj}), and (\ref{chij}), we introduce $\hat{C}({\bf r}_j)$, $\hat{C}^{\dagger}({\bf r}_j)$, $\hat{N}({\bf r}_j)$, and $\hat{\chi}({\bf r}_j)$,
 \begin{eqnarray}
\hat{C}({\bf r}_j)=\int_{V_j} d{\bf r} \hat{\psi}_{e}({\bf r})=e^{- { i \over 2} \hat{\chi}({\bf r}_j)} \hat{N}({\bf r}_j)^{ 1 \over 2}; \quad  
\hat{C}^{\dagger}({\bf r}_j)=\int_{V_j} d{\bf r} \hat{\psi}_{e}({\bf r})=e^{ { i \over 2} \hat{\chi}({\bf r}_j)} \hat{N}({\bf r}_j)^{ 1 \over 2}
 \end{eqnarray}
 where $V_j$ is the $j$th coarse-grained cell.
 
 The eigenvalue of the number operator $\hat{N}({\bf r}_j)$ can be interpreted as the number of electrons in the collective mode for the supercurrent in the $j$th cell. The phase factor operators $e^{\pm { i \over 2}\hat{\chi}({\bf r}_j)}$ change the eigenstate as
  \begin{eqnarray}
e^{\pm { i \over 2}\hat{\chi}({\bf r}_j)}|N({\bf r}_j)\rangle =|N({\bf r}_j) \pm 1\rangle 
 \end{eqnarray}
 
 Using $e^{\pm { i }\hat{\chi}({\bf r}_j)}$, the interaction part of the Hamiltonian at the $j$th cell can be written as
\begin{eqnarray}
H_{\rm int}= \sum_{{\bf k} {\bm \ell}} V_{{\bf k} {\bm \ell}}c^{\dagger}_{{\bf k} \uparrow}c^{\dagger}_{-{\bf k} \downarrow} e^{- { i }\hat{\chi}({\bf r}_j)}   e^{{ i }\hat{\chi}({\bf r}_j)}c_{-{\bm \ell} \downarrow}c_{{\bm \ell} \uparrow}
\end{eqnarray}
which can be transformed to a mean-field version
\begin{eqnarray}
H^{\rm MF}_{\rm int}&=& \sum_{{\bf k} {\bm \ell}} V_{{\bf k} {\bm \ell}}
\Big[ \langle c^{\dagger}_{{\bf k} \uparrow}c^{\dagger}_{-{\bf k} \downarrow} e^{- { i }\hat{\chi}({\bf r}_j)}    \rangle
e^{{ i }\hat{\chi}({\bf r}_j)}c_{-{\bm \ell} \downarrow}c_{{\bm \ell} \uparrow}+
c^{\dagger}_{{\bf k} \uparrow}c^{\dagger}_{-{\bf k} \downarrow} e^{- { i }\hat{\chi}({\bf r}_j)}   
\langle e^{{ i }\hat{\chi}({\bf r}_j)}c_{-{\bm \ell} \downarrow}c_{{\bm \ell} \uparrow} \rangle
\nonumber
\\
&-& \langle
c^{\dagger}_{{\bf k} \uparrow}c^{\dagger}_{-{\bf k} \downarrow} e^{- { i \over 2}\hat{\chi}({\bf r}_j)} \rangle
\langle
  e^{{ i \over 2}\hat{\chi}({\bf r}_j)}c_{-{\bm \ell} \downarrow}c_{{\bm \ell} \uparrow} \rangle
\Big]
\nonumber
\\
&=& \!-\! \sum_{{\bf k}} \Big[ 
\Delta_{\bf k} ({\bf r}_j)c^{\dagger}_{{\bf k} \uparrow}c^{\dagger}_{-{\bf k} \downarrow} e^{- { i }\hat{\chi}({\bf r}_j)}   
\!+\! \Delta^{\ast}_{\bf k} ({\bf r}_j)e^{{ i }\hat{\chi}({\bf r}_j)}c_{-{\bf k} \downarrow}c_{{\bf k} \uparrow}
\!-\! \Delta_{\bf k} ({\bf r}_j) \langle
c^{\dagger}_{{\bf k} \uparrow}c^{\dagger}_{-{\bf k} \downarrow} e^{- { i \over 2}\hat{\chi}({\bf r}_j)} \rangle
\Big]
\label{MF}
\end{eqnarray}
where
\begin{eqnarray}
\Delta_{\bf k} ({\bf r}_j)=-\sum_{{\bm \ell}} V_{{\bf k} {\bm \ell}}\langle e^{{ i }\hat{\chi}({\bf r}_j)}c_{-{\bf k} \downarrow}c_{{\bf k} \uparrow} \rangle
\end{eqnarray}

Note that the expectation values used to obtained the mean-field Hamiltonian can be calculated with a particle number conserved state as in the usual Hartree-Fock method since the operators $c^{\dagger}_{{\bf k} \uparrow}c^{\dagger}_{-{\bf k} \downarrow} e^{- { i \over 2}\hat{\chi}({\bf r}_j)}$ and  
$e^{{ i \over 2}\hat{\chi}({\bf r}_j)}c_{-{\bm \ell} \downarrow}c_{{\bm \ell} \uparrow}$ conserve the number of electrons. This is a marked contrast to the standard theory in which the expectation values are calculated for $c^{\dagger}_{{\bf k} \uparrow}c^{\dagger}_{-{\bf k} \downarrow}$ and  
$c_{-{\bm \ell} \downarrow}c_{{\bm \ell} \uparrow}$ that do not conserve the number of electrons.

If we replace the operators $e^{\pm { i \over 2}\hat{\chi}({\bf r}_j)}$ by their eigenvalues $e^{\pm { i \over 2} {\chi}({\bf r}_j)}$ and calculate the expectation values using the BCS state vector, we have
\begin{eqnarray}
 \Delta^{\rm BCS}_{\bf k} ({\bf r}_j)=-\sum_{{\bm \ell}} V_{{\bf k} {\bm \ell}} \sin \theta_{\bm \ell} ({\bf r}_j) \cos\theta_{\bm \ell} ({\bf r}_j)
\end{eqnarray}
This is the formula for the energy gap in the standard theory of superconductivity.

We may use the following modified Bogoliubov transformation,
\begin{eqnarray}
\gamma_{{\bf k} 0}({\bf r}_j) &=&u_{\bf k} ({\bf r}_j) c_{{\bf k} \uparrow}- v_{\bf k} ({\bf r}_j) c^{\dagger}_{-{\bf k} \downarrow} e^{- i\hat{\chi}({\bf r}_j)}
\nonumber
\\
\gamma_{{\bf k} 1}({\bf r}_j) &=& u_{\bf k} ({\bf r}_j) c_{-{\bf k} \downarrow}+ v_{\bf k} ({\bf r}_j) c^{\dagger}_{{\bf k} \uparrow} e^{- i\hat{\chi}({\bf r}_j)}
\end{eqnarray}
where the operator $u_{\bf k} ({\bf r}_j)$ $v_{\bf k} ({\bf r}_j)$ are real parameters that satisfy $u^2_{\bf k} ({\bf r}_j)+v^2_{\bf k} ({\bf r}_j)=1$, and $e^{- i\hat{\chi}({\bf r}_j)}$ is the operator that annihilate two electrons. Such an operator was introduced previously \cite{Josephson62,TinkhamText,Bardeen62,Tinkham72}; however, they are absent in the standard theory now.

Using the above operators, and assuming that $e^{- i\hat{\chi}({\bf r}_j)}$ commute with $c_{{\bf k} \sigma}$ and
$c^{\dagger}_{{\bf k} \sigma}$
the Hamiltonian in the $j$th cell is cast in the form
\begin{eqnarray}
H^{\rm MF}({\bf r}_j) &=& \sum_{\bf k}E_{\bf k}({\bf r}_j)[ \gamma^{\dagger}_{{\bf k} 0}({\bf r}_j)  \gamma_{{\bf k} 0}({\bf r}_j) +\gamma^{\dagger}_{{\bf k} 1}({\bf r}_j)\gamma_{{\bf k} 1}({\bf r}_j)]
\nonumber
\\
&+& \sum_{\bf k}\left(\xi_0({\bf k}) -  E_{\bf k}({\bf r}_j)+ \Delta_{\bf k} ({\bf r}_j) \langle
c^{\dagger}_{{\bf k} \uparrow}c^{\dagger}_{-{\bf k} \downarrow} e^{- { i \over 2}\hat{\chi}({\bf r}_j)} \rangle \right) 
\end{eqnarray}
where $E_{\bf k}({\bf r}_j)$ is the Bogoliubov quasi-particle energy;
$E_{\bf k}({\bf r}_j)$, $u_{\bf k} ({\bf r}_j)$,$v_{\bf k} ({\bf r}_j)$, and $\Delta_{\bf k} ({\bf r}_j)$ are self-consistently obtained from the relations,
\begin{eqnarray}
&&E_{\bf k}({\bf r}_j) = [ \Delta^2_{\bf k} ({\bf r}_j)+ \xi^2_0({\bf k})]^{1/2},
\quad u^2_{\bf k} ({\bf r}_j)={ 1 \over 2} \left( 1+{ { \xi_0({\bf k})} \over {E_{\bf k}({\bf r}_j) }}\right)
\nonumber
\\
&&
v^2_{\bf k} ({\bf r}_j)=1-u^2_{\bf k} ({\bf r}_j), \quad \Delta_{\bf k} ({\bf r}_j)=- \sum_{\bm \ell } V_{{\bf k} {\bm \ell}}
 u_{\bf k} ({\bf r}_j)v_{\bf k} ({\bf r}_j)
\end{eqnarray}

The ground state is the vacuum of $\gamma_{{\bf k} 0}({\bf r}_j)$ and $\gamma_{{\bf k} 1}({\bf r}_j)$. It is given by
\begin{eqnarray}
|g ({\bf r}_j,t) \rangle=\prod_{\bf k} \left( u_{\bf k} ({\bf r}_j) +  v_{\bf k} e^{-i \hat{\chi}({\bf r}_j,t) } ({\bf r}_j) c^{\dagger}_{{\bf k} \uparrow}c^{\dagger}_{-{\bf k} \downarrow} \right) |{\rm cnd({\bf r}_j)} \rangle
\end{eqnarray}
where $|{\rm cnd({\bf r}_j)} \rangle$ is the state vector for the condensate state that has $N({\bf r}_j)$ electrons in the collective mode described by $\chi$.
We may construct $|{\rm cnd({\bf r}_j)} \rangle$ from $\Psi$ in Eq.~(\ref{f}): $N({\bf r}_j)$ is identified as the number of electrons in the $j$th cell calculated 
with $\Psi$;  $\chi({\bf r}_j)$ may be obtained as the value of $\chi({\bf r})$ at the center of the $j$th cell, ${\bf r}_j$. Note that $\chi({\bf r}_j)$'s are not physically meaningful values, but phase differences $\chi({\bf r}_j)-\chi({\bf r}_k)$'s between nearby cells are.

The original BCS formulae are obtained by replacing $\hat{\chi}({\bf r}_j)$ with a scalar ${\chi}({\bf r}_j)$, and $|{\rm cnd({\bf r}_j)} \rangle$ by $|{\rm vac} \rangle$.
In the new theory, the existence of $|{\rm cnd({\bf r}_j)} \rangle$ is needed prior to the electron-pairing gap formation to have superconducting states.
 In other words, the origin of $\chi$ must be sought separately from identifying the interaction for the energy gap formation.
 
 \section{Wave-Packet Dynamics of Bloch Electrons in the Presence of Rashba Spin-Orbit Interaction and Magnetic Field}
 \label{section5}

The normal state of the BCS superconductors is a band metal. It exhibits quantum oscillations when a magnetic field is applied.
This oscillation is due to the reorganization of electronic states near the Fermi surface.
In this section, we examine this reorganization in the presence of the weak Rashba spin-orbit coupling compared to the electron-pairing energy gap.

In order to include the effect of the magnetic field ${\bf B}^{\rm em}=\nabla \times {\bf A}^{\rm em}$ that gives rise to the cyclotron motion, we use the wave-packet dynamics formalism \cite{Niu}.
We consider electrons in a single band and denote its Bloch wave as
\begin{eqnarray}
|\psi_{\bf q}\rangle = e^{i {\bf q} \cdot {\bf r}} |u_{\bf q}\rangle
\end{eqnarray}
where ${\bf q}$ is the wave vector and $|u_{\bf q}\rangle$ is the periodic part of the Bloch wave.

It satisfies the Schr\"{o}dinger equation,
\begin{eqnarray}
H_0[{\bf q}]|u_{\bf q}\rangle ={\cal E}({\bf q}) |u_{\bf q}\rangle,
\end{eqnarray}
where $H_0$ is the zeroth order single-particle Hamiltonian for an electron in a  periodic potential.

According to the wave packet dynamics formalism, $H_0[{\bf q}]$ is modified as
\begin{eqnarray}
H_0[{\bf q}] \rightarrow H_0 \left[{\bf q}+{ e \over {\hbar}} {\bf A}^{\rm em}({\bf r})\right].
\end{eqnarray}
in the presence of the magnetic field ${\bf B}^{\rm em}=\nabla \times {\bf A}^{\rm em}$.

Using the Bloch waves, a wave-packet centered at coordinate ${\bf r}_c$ and with central wave vector ${\bf q}_c$ is constructed as
\begin{eqnarray}
\langle {\bf r}|({\bf q}_c, {\bf r}_c) \rangle &=& \int d^3q \ a({\bf q},t) 
\langle {\bf r}|\psi_{ {\bf q}}\rangle 
 e^{-i {1 \over 2} \chi({\bf r})}\left(
 \begin{array}{c}
 e^{i { 1 \over 2} \xi ({\bf r})} \sin {{\zeta ({\bf r}) } \over 2}
 \\
  e^{-i { 1 \over 2} \xi {\bf r})} \cos {{\zeta ({\bf r}) } \over 2} 
 \end{array}
 \right)
\label{wp}
\end{eqnarray}
where $a({\bf q})$ is a distribution function, and the spin function is the one given in Eq.~(\ref{spin-direction}). The wave packet with the spin function in  Eq.~(\ref{spin-direction2}) can be constructed, analogously.

The distribution function $a({\bf q},t)$ satisfies the normalization 
\begin{eqnarray}
 \int d^3{q} \, |a({\bf q},t)|^2&=&1
 \end{eqnarray}
 and the localization condition in ${\bf k}$ space,
 \begin{eqnarray}
 \int d^3{q} \, {\bf q} |a({\bf q},t)|^2&=&{\bf q}_c
\end{eqnarray}
The distribution of $|a({\bf q},t)|^2$ is assumed to be narrow compared with the Brillouin zone size so that ${\bf q}_c$ can be regarded as the central wave vector of the wave packet.

The wave packet is also localized in ${\bf r}$ space around the central position ${\bf r}_c$,
\begin{eqnarray}
{\bf r}_c&=& \langle ({\bf q}_c, {\bf r}_c) |{\bf r} |({\bf q}_c, {\bf r}_c) \rangle.
\end{eqnarray}

The crucial ingredient for realizing the spin-twisting itinerant motion is the Rashba spin-orbit interaction. We include the following Rashba interaction term in the Hamiltonian
\begin{eqnarray}
H_{so}= {\bm \lambda}({\bf r})\cdot {{\hbar{\bm \sigma}} \over 2} \times \left(\hat{\bf p}-{q}{\bf A}^{\rm em}({\bf r})\right), 
\end{eqnarray}
where ${\bm \lambda}({\bf r})$ is the spin-orbit coupling vector (its direction is the internal electric field direction), ${\bf r}$ is the spatial coordinates, $\hat{\bf p}=-i\hbar \nabla$ is the momentum operator, and $q=-e$ is electron charge \cite{Rashba}. 

Let us construct the Lagrangian $L'({\bf r}_c, \dot{\bf r}_c,{\bf q}_c,\dot{\bf q}_c)$ using the time-dependent variational principle, 
\begin{eqnarray}
L'=\langle ({\bf q}_c, {\bf r}_c) | i \hbar {\partial \over {\partial t}}-H| ({\bf q}_c, {\bf r}_c) \rangle.
\end{eqnarray}

For convenience sake, we introduce another Lagrangian $L$ that is related to $L'$ as 
\begin{eqnarray}
L=L'-\hbar{d \over {dt}}\left[ \gamma({\bf q}_c,t) -{\bf r}_c \cdot {\bf q}_c \right],
\end{eqnarray}
where $\gamma$ is the phase of $a({\bf q},t)=|a({\bf q},t)|e^{-i \gamma({\bf q},t)}$.

By following  procedures for calculating expectation values for operators by the wave packet \cite{Niu}, 
$L$ is obtained as 
\begin{eqnarray}
L&=&-{\cal E}\left({\bf q}_c+{ e \over {\hbar}} {\bf A}^{\rm eff}({\bf r}_c)\right) 
+\hbar {\bf q}_c \cdot \dot{{\bf r}_c}+i\hbar \left\langle u_{\bf q} \left| {{d u_{\bf q} } \over {dt } } \right. \right\rangle
\nonumber
\\
&+&\hbar {\bm \lambda}({\bf r}_c)\cdot \left[{\bf s({\bf r}_c)} \times \left({\bf q}_c+{ e \over {\hbar}} {\bf A}^{\rm eff}({\bf r}_c) \right)\right],
\end{eqnarray}
 where ${\bf s}({\bf r}_c)$ is the expectation value of spin for the wave packet centered at ${\bf r}_c$ given by
 \begin{eqnarray}
{\bf s}({\bf r}_c)={\hbar \over 2} \langle ({\bf q}_c, {\bf r}_c) | {\bm \sigma} |({\bf q}_c, {\bf r}_c)\rangle.
\end{eqnarray}

We introduce the following wave vector ${\bf k}_c$, 
\begin{eqnarray}
{\bf k}_c = {\bf q}_c+{ e \over {\hbar}} {\bf A}^{\rm eff}({\bf r}_c)
\label{gaugek}
\end{eqnarray}
and change the dynamical variables from ${\bf q}_c, \dot{\bf q}_c$ to ${\bf k}_c, \dot{\bf k}_c$ \cite{Niu}.
 
Then, the Lagrangian with dynamical variables ${\bf r}_c, \dot{\bf r}_c,{\bf k}_c,\dot{\bf k}_c$ is given by
\begin{eqnarray}
&&L({\bf r}_c, \dot{\bf r}_c,{\bf k}_c,\dot{\bf k}_c)=-{\cal E}({\bf k}_c) +\hbar {\bm \lambda}({\bf r}_c)\cdot \left[{\bf s}({\bf r}_c) \times {\bf k}_c \right]
\nonumber
\\
&+&\hbar \left[ {\bf k}_c -{ e \over {\hbar}} {\bf A}^{\rm eff}({\bf r}_c) \right] \cdot \dot{{\bf r}_c}
+i\hbar \dot{{\bf k}}_c \cdot \left \langle u_{\bf q} | {{\partial u_{\bf q} } \over {\partial {\bf q} } } \right\rangle_{ {\bf q}={\bf k}_c}
\label{Lag}
\end{eqnarray}

Using the above Lagrangian $L$, the following equations of motion are obtained,
\begin{eqnarray}
\dot{\bf r}_c&=&{ 1 \over \hbar} {{\partial {\cal E}} \over {\partial {\bf k}_c}}+ {\bm \lambda}({\bf r}_c) \times {\bf s}({\bf r}_c)-\dot{\bf k}_c\times {\bm \Omega},
\label{eqm1}
\\
\dot{\bf k}_c&=&   { {  \partial  } \over {\partial {\bf r}_c}}\left[{\bm \lambda}({\bf r}_c) \times{\bf s}({\bf r}_c) \cdot {\bf k}_c  \right] -{e \over {\hbar }}\dot{\bf r}_c\times {\bf B}^{\rm eff},
\label{eqm2}
\end{eqnarray}
where ${\bm \Omega}$ is the Berry curvature in ${\bf k}$ space defined by
\begin{eqnarray}
{\bm \Omega}=i\hbar\nabla_{\bf q} \times  \left\langle u_{\bf q} | \nabla_{\bf q}| u_{\bf q} \right\rangle
\end{eqnarray}
and ${\bf B}^{\rm eff}$ is the effective magnetic field,
\begin{eqnarray}
{\bf B}^{\rm eff}=\nabla \times {\bf A}^{\rm eff}={\bf B}^{\rm em}+{\hbar \over {2q}} \nabla \times \nabla \chi
\end{eqnarray}

In the following, we consider the case where ${\bm \Omega}=0$. 
Then, Eq.~(\ref{eqm1}) becomes
\begin{eqnarray}
\dot{\bf r}_c={ 1 \over \hbar} {{\partial {\cal E}({\bf k}_c)} \over {\partial {\bf k}_c}}+{\bm \lambda}({\bf r}_c)\times {\bf s}({\bf r}_c).
\label{eqm3}
\end{eqnarray}

Using Eq.~(\ref{eqm3}), and (\ref{eqm2}) becomes,
\begin{eqnarray}
\dot{\bf k}_c &=&  { {  \partial  } \over {\partial {\bf r}_c}}\left[\left(\dot{\bf r}_c-{ 1 \over \hbar} {{\partial {\cal E}({\bf k}_c)} \over {\partial {\bf k}_c}}\right) \cdot {\bf k}_c  \right] -{e \over {\hbar}}\dot{\bf r}_c\times {\bf B}^{\rm eff}
\nonumber
\\
&=&-{e \over {\hbar}}\dot{\bf r}_c\times {\bf B}^{\rm eff}
\label{eqm4}
\end{eqnarray}

 Eqs.~(\ref{eqm3}) and (\ref{eqm4}) indicate that the wave packet exhibits cyclotron motion for the electron in the band with energy
\begin{eqnarray}
{\cal E}({\bf k})+\hbar{\bm \lambda}({\bf r})\times {\bf s}({\bf r})\cdot {\bf k}
\end{eqnarray}

By following the Onsager's argument, let us quantize the cyclotron orbit \cite{Onsager1952}. From Eq.~(\ref{Lag}), the Bohr-Sommerfeld relation becomes
\begin{eqnarray}
\oint_C (\hbar {\bf k}_c -e{\bf A}^{\rm eff}) \cdot d{\bf r}_c =2\pi \hbar \left(n+{ 1 \over 2} \right)
\label{Onsager1}
\end{eqnarray}
where $n$ is an integer and $C$ is the closed loop that corresponds to the section of Fermi surface enclosed by the cyclotron motion.

From Eq.~(\ref{eqm4}), we have 
\begin{eqnarray}
\oint_C \hbar {\bf k}_c \cdot d{\bf r}_c=-e \oint_C d{\bf r}_c \cdot {\bf r}_c \times {\bf B}^{\rm eff}=e \oint_C  {\bf B}^{\rm eff} \cdot  {\bf r}_c \times d{\bf r}_c
\end{eqnarray}

We consider the situation where a singularity of $\chi$ exists within $C$, and the magnetic field ${\bf B}^{\rm em}$ is uniform. 
Then, the above equation becomes
\begin{eqnarray}
\oint_C \hbar {\bf k}_c \cdot d{\bf r}_c=e {\bf B}^{\rm em} \cdot \oint_C {\bf r}_c \times d{\bf r}_c  =2e \oint_C  {\bf A}^{\rm em} \cdot d {\bf r}_c 
\end{eqnarray}

Thus, the l.h.s. of Eq.~(\ref{Onsager1}) is calculated as
\begin{eqnarray}
2e \oint_C  {\bf A}^{\rm em} \cdot  d{\bf r}_c-e \oint_C  {\bf A}^{\rm em} \cdot  {\bf r}_c +{ \hbar \over 2} \oint_C \nabla_{{\bf r}_c} \chi \cdot d {\bf r}_c
=e \oint_C  {\bf A}^{\rm em} \cdot  d{\bf r}_c+{ \hbar \pi}w_C[\chi]
\label{Onsager2}
\end{eqnarray}

This leads to the quantization of the cyclotron motion given by
\begin{eqnarray}
e \oint_C  {\bf A}^{\rm em} \cdot  d{\bf r}_c+{ \hbar \pi}w_C[\chi]=2\pi \hbar \left(n+{ 1 \over 2} \right)
\label{Onsager3}
\end{eqnarray}

The important point is that above condition is satisfied even the magnetic field is absent. In this case, the first term in the l.h.s. is zero; still, the relation holds for $w_C[\chi]=1, n=0$ and $w_C[\chi]=-1, n=-1$. This will be interpreted that the $\pi$-flux Dirac string provides a magnetic flux for the zero-point cyclotron motion.

\section{The pairing energy gap}
\label{section6}

Instead of the pairing between single particle states $({\bf k}, \uparrow)$ and $(-{\bf k}, \downarrow)$, we consider the pairing between $({\bf k}_c, {\bf s}_0({\bf r}_c))$ and $(-{\bf k}_c, -{\bf s}_0({\bf r}_c))$. 
We will obtain the pairing energy gap at ${\bf r}_c$ by treating the wave packets (${\bf k}_c, {\bf r}_c$) as basis states in each corse-gained cell centered at ${\bf r}_c$.

The single-particle energy for the states $({\bf k}_c, {\bf s}_0({\bf r}_c))$ and $(-{\bf k}_c, -{\bf s}_0({\bf r}_c))$ are given by
\begin{eqnarray}
  {\cal E}_{+}({\bf k}_c, {\bf r}_c)={\cal E}({\bf k}_c)+\hbar {\bm \lambda} ({\bf r}_c) \times {\bf k}_c \cdot {\bf s}_0({\bf r}_c)
\end{eqnarray}
where ${\cal E}({\bf k}_c)={\cal E}(-{\bf k}_c)$ is assumed.

Another pairing of states $({\bf k}_c, -{\bf s}_0({\bf r}_c))$ and $(-{\bf k}_c, {\bf s}_0({\bf r}_c))$ are possible. Their single-particle energy is
   \begin{eqnarray}
      {\cal E}_{-}({\bf k}_c, {\bf r}_c)=   {\cal E}({\bf k}_c)- \hbar {\bm \lambda}({\bf r}_c) \times {\bf k}_c\cdot {\bf s}_0({\bf r}_c)
   \label{e2}
 \end{eqnarray}

Now, we come back to the pairing of $({\bf k}_c, {\bf s}_0({\bf r}_c))$ and $(-{\bf k}_c, -{\bf s}_0({\bf r}_c))$, and also  $({\bf k}_c, -{\bf s}_0({\bf r}_c))$ and $(-{\bf k}_c, {\bf s}_0({\bf r}_c))$. The parameters for the pairing and energy gap are now functions of ${\bf k}_c$ and ${\bf r}_c$; $u_{\bf k}$ and $v_{\bf k}$ are replaced by $u_{\pm}({\bf k}_c, {\bf r}_c)$ and $v_{\pm}({\bf k}_c, {\bf r}_c)$ given by
\begin{eqnarray}
u_{\pm}({\bf k}_c, {\bf r}_c)&=&{1 \over \sqrt{2}} \left(1 + {{{\xi}_{\pm}({\bf k}_c, {\bf r}_c)} \over \sqrt{{\xi}^2_{\pm}({\bf k}_c, {\bf r}_c)+ \Delta^2( {\bf r}_c)}} \right)^{1/2},
\nonumber
\\
&&
\\
v_{\pm}({\bf k}_c, {\bf r}_c)&=&{1 \over \sqrt{2}} \left(1 - {{{\xi}_{\pm}({\bf k}_c, {\bf r}_c)} \over \sqrt{{\xi}^2_{\pm}({\bf k}_c, {\bf r}_c)+ \Delta^2 ({\bf r}_c)}} \right)^{1/2},
\nonumber
\\
\end{eqnarray}
where
\begin{eqnarray}
\xi_{\pm}({\bf k})={\cal E}_{\pm}({\bf k})-{\cal E}_{F}={\cal \xi}_0({\bf k}_c)\pm\hbar {\bm \lambda} ({\bf r}_c) \times {\bf k}_c \cdot {\bf s}_0({\bf r}_c)
\end{eqnarray}
and the gap function $\Delta( {\bf r}_c)$ is the solution of the gap equation given by
\begin{eqnarray}
\Delta({\bf r}_c)\!&=&\!{{g} \over 2}\sum_{|{\xi}_{0}({\bm \ell})| < \hbar \omega_D }  \left\{ u_{+}({\bm \ell }_c, {\bf r}_c)v_{+}({\bm \ell}_c, {\bf r}_c)
\!+\!u_{-}({\bm \ell}_c, {\bf r}_c)v_{-}({\bm \ell}_c, {\bf r}_c) \right\}
\nonumber
\\
\!&=&{{g \Delta({\bf r}_c)} \over 4}\!\sum_{|{\xi}_{0}({\bm \ell})| < \hbar \omega_D }  \left\{ {{1} \over \sqrt{{\xi}^2_{+}({\bf k}_c, {\bf r}_c)+ \Delta^2 ({\bf r}_c)}} 
\!+\! {{1} \over \sqrt{{\xi}^2_{-}({\bf k}_c, {\bf r}_c)+ \Delta^2 ({\bf r}_c)}}   \right\}
\nonumber
\\
\!&\approx&{{g \Delta({\bf r}_c)} \over 4}\!\sum_{|{\xi}_{0}({\bm \ell})| < \hbar \omega_D }  \left\{ {{2} \over \sqrt{{\xi}^2_{0}({\bf k}_c, {\bf r}_c)+ \Delta^2 ({\bf r}_c)}} 
\!-\! {{\lambda^2} \over {[{\xi}^2_{0}({\bf k}_c, {\bf r}_c)+ \Delta^2 ({\bf r}_c)]^{3/2}}}   \right\}
\nonumber
\\
\!&\approx&{{g \Delta({\bf r}_c) N(0;{\bf r}_c)} \over 4} \int_{-\hbar \omega_D}^{\hbar \omega_D} d \epsilon  \left\{ {{2} \over \sqrt{\epsilon^2+ \Delta^2 ({\bf r}_c)}} 
\!-\! {{\lambda^2} \over {[{\epsilon}^2+ \Delta^2 ({\bf r}_c)]^{3/2}}}   \right\}
\nonumber
\\
\!&\approx&{{g \Delta({\bf r}_c) N(0;{\bf r}_c)}} \left\{ \log {{2 \hbar \omega_D} \over {\Delta}} 
\!-\! {{\lambda^2} \over {\Delta^2}}   \right\}
\label{gap}
\end{eqnarray}
where $N(0;{\bf r}_c)$ is the density of states at the Fermi energy in the corse grained cell of center ${\bf r}_c$.

From the above relation, we have 
\begin{eqnarray}
\Delta({\bf r}_c) \approx  2 \hbar \omega_D \exp \left( - { 1 \over {g N(0;{\bf r}_c)}} -{ {\lambda^2}  \over {\Delta_0^2}} \right ); \quad \Delta_0({\bf r}_c) = 2 \hbar 
\omega_D \exp \left( - { 1 \over {g N(0;{\bf r}_c)}} \right)
 \label{gap2}
 \end{eqnarray}
 where we assume that $\hbar \omega_D \gg \Delta$. 
 The gap $\Delta$ is reduced by the spin-orbit interaction, generally. If the spin-orbit interaction parameter $\lambda$ is significantly smaller that $\Delta_0$, the gap becomes the original one.

\section{The Kinetic Energy with Rashba Interaction and London Equation}
\label{section8}

The kinetic energy density including the Rashba interaction is given by
\begin{eqnarray}
 2\sum_{\bf k} {\xi}_-({\bf k}, {\bf r})  v^2_-({\bf k}, {\bf r})+2\sum_{\bf k} {\xi}_+({\bf k}, {\bf r})  v^2_+({\bf k}, {\bf r}) 
 \label{Ekin}
\end{eqnarray}

For simplicity, we approximate it using the Fermi distribution functions $f(\epsilon)=(1+ e^{\epsilon/k_BT})^{-1}$ ($k_B$ is Boltzmann's constant) and density of states $N(\epsilon;{\bf r}_c)$ as
\begin{eqnarray}
&&\int  {{N(\epsilon;{\bf r}_c)} \over 2} \Big\{ [\epsilon+ \hbar {\bm \lambda}({\bf r}_c) \times {\bf k}_c\cdot {\bf s}_0({\bf r}_c)]
f(\epsilon+ \hbar {\bm \lambda}({\bf r}_c) \times {\bf k}_c\cdot {\bf s}_0({\bf r}_c))
\nonumber
\\
&+&
[\epsilon- \hbar {\bm \lambda}({\bf r}_c) \times {\bf k}_c\cdot {\bf s}_0({\bf r}_c)]
f(\epsilon-\hbar {\bm \lambda}({\bf r}_c) \times {\bf k}_c\cdot {\bf s}_0({\bf r}_c))
 \Big\}d \epsilon
\nonumber
 \\
  &\approx&
  \int  {{N(\epsilon;{\bf r}_c)} \over 2} \Big\{ \epsilon \left[f(\epsilon+ \hbar {\bm \lambda}({\bf r}_c) \times {\bf k}_c\cdot {\bf s}_0({\bf r}_c))+f(\epsilon- \hbar {\bm \lambda}({\bf r}_c) \times {\bf k}_c\cdot {\bf s}_0({\bf r}_c)) \right]
  \nonumber
\\
  &+& \hbar {\bm \lambda}({\bf r}_c) \times {\bf k}_c\cdot {\bf s}_0({\bf r}_c) \left[f(\epsilon+ \hbar {\bm \lambda}({\bf r}_c) \times {\bf k}_c\cdot {\bf s}_0({\bf r}_c))-f(\epsilon- \hbar {\bm \lambda}({\bf r}_c) \times {\bf k}_c\cdot {\bf s}_0({\bf r}_c)) \right] \Big\} d \epsilon
\nonumber
\\
 &\approx &
  \int  {{N(\epsilon;{\bf r}_c)} \over 2} \left\{ 2 \epsilon f(\epsilon) 
  + 2\left|\hbar {\bm \lambda}({\bf r}_c) \times {\bf k}_c\cdot {\bf s}_0({\bf r}_c) \right|^2 {{\partial f(\epsilon)} \over {\partial \epsilon}}  \right\}d \epsilon
\label{energy0}
\end{eqnarray}

At temperature $T=0$,  ${{\partial f(\epsilon)} \over {\partial \epsilon}} =- \delta( \epsilon)$; thus,  the above becomes, 
\begin{eqnarray}
  \int d \epsilon N(\epsilon;{\bf r}_c)  \epsilon f(\epsilon) d \epsilon- N(0;{\bf r}_c)  \left|\hbar {\bm \lambda}({\bf r}_c) \times {\bf k}_c\cdot {\bf s}_0({\bf r}_c) \right|^2 
  \label{energy0a}
\end{eqnarray}

The first term may be approximated as
\begin{eqnarray}
  \int d \epsilon N(\epsilon;{\bf r}_c)  \epsilon f(\epsilon)
  \approx \sum_{\xi_0({\bf q}) <0}
  { {\hbar^2} \over {2m}}\left[ {\bf q}+{e \over \hbar } {\bf A}^{\rm eff} \right]^2
  \approx
  \sum_{q < q_{F}}
  { {\hbar^2} \over {2m}}{\bf q}^2+{{e^2 \rho({\bf r}_c)} \over {2m}} |{\bf A}^{\rm eff}|^2 
  \label{energy0a}
\end{eqnarray}
assuming that the term linear in ${\bf q}$ cancels out.

The second term may be approximated as
\begin{eqnarray}
&&- N(0;{\bf r}_c)  \left|\hbar {\bm \lambda}({\bf r}_c) \times {\bf k}_c\cdot {\bf s}_0({\bf r}_c) \right|^2 
    \approx  - \sum_{\xi_0({\bf q}) =0}\left|\hbar {\bm \lambda}({\bf r}_c) \times \left[ {\bf q}+{e \over \hbar } {\bf A}^{\rm eff} \right]\cdot {\bf s}_0({\bf r}_c) \right|^2 
    \nonumber
    \\
     &\approx&  - \hbar^2\sum_{\xi_0({\bf q}) =0}\left| {\bm \lambda}({\bf r}_c) \times {\bf q}\cdot {\bf s}_0({\bf r}_c) \right|^2 
      - e^2 N(0;{\bf r}_c)\left| {\bm \lambda}({\bf r}_c) \times {\bf s}_0({\bf r}_c) \cdot {\bf A}^{\rm eff}\right|^2   
  \label{energy0b}
\end{eqnarray}
assuming that the term linear in ${\bf q}$ cancels out.

To minimize the kinetic energy, ${\bf s}_0$ is so chosen to satisfy 
\begin{eqnarray}
{\bm \lambda ({\bf r})} \times {\bf s}_0({\bf r})  \parallel {\bf A}^{\rm eff}({\bf r}) 
\label{conds0}
\end{eqnarray}

Then, the current density is given by
\begin{eqnarray}
{\bf j}_{\rm tot}({\bf r})=- e^2  \left[{{\rho ({\bf r})} \over {m}} - N(0; {\bf r}) |{\bm \lambda} ({\bf r})\times {\bf s}_0 ({\bf r})|^2 \right]
{\bf A}^{\rm eff}({\bf r})
\label{current3}
\end{eqnarray}
where the contribution from the energy gap term is neglected by assuming it is very small.
This is the London equation, and the system should exhibit the Meissner effect. 

 When the magnetic field is absent we replace ${\bf A}^{\rm eff}$ by ${\hbar \over {2q}} \nabla \chi$. Then, the kinetic energy increase given in Eq.~(\ref{energy0a}) is calculated as (taking the volume of the coarse-grained cell unity)
 \begin{eqnarray}
\int d^3 r  {{e^2 \rho({\bf r})} \over {2m}} |{\bf A}^{\rm eff}|^2 \approx { \hbar^2 \over {8m}}\rho_0 \int d^3 r (\nabla \chi)^2={ {\hbar^2 \rho_0} \over {8m}} \int_{Surface} 
d{\bf S} \cdot ( \chi \nabla \chi)
 \end{eqnarray}
 where we assume that $\rho$ is constant in the bulk ($\rho=\rho_0$), and the relation $\nabla^2 \chi=0$ is used. This surface term is negligibly small compared to the bulk energy if the system is sufficiently large. The energy gain in Eq.~(\ref{energy0b}) is in the order of $\lambda^2$ and the energy deficit from the decrease of the gap  in Eq.~(\ref{gap2}) is in the order of $ e^{ -\lambda^2 \Delta_0^{-2}}$, thus, the system gain energy by changing the electron pairing states. Actually, the creation of the lines of singularities costs the core energies. Therefore, the density of them will be determined by the competition between the energy gain by Eq.~(\ref{energy0b}) and the energy cost by the creation of the singularities.
 
\section{Critical look at the gauge invariance problem in the BCS theory}
\label{section4}

In the original BCS calculation, the Meissner effect is explained as a linear response to an applied magnetic field by treating ${\bf A}^{\rm em}\neq 0$ as a perturbation for the wave function obtained for the gauge ${\bf A}^{\rm em} =0$ \cite{BCS1957}. 
 
The BCS employed the following gauge,
\begin{eqnarray}
\nabla \cdot {\bf A}^{\rm em}=0; \quad {\bf A}^{\rm em}=0 \  \mbox{ if the magnetic field is zero.}
\label{BCSa}
\end{eqnarray}

The obtained current was not gauge invariant, and the validity of using the gauge $\nabla \cdot {\bf A}^{\rm em}=0$ was intensively studied by a number of researchers \cite{Nambu1960,Buckingam1957,Schafroth1958,Anderson1958a,Anderson1958b,Rickayzen1958,Yoshida1959}, and believed to be solved.
Most notably, Nambu using the Ward-Takahashi identity \cite{Nambu1960} performed the gauge invariant Meissner effect calculation. This lead to discover the collective mode of paired-electrons that restores the gauge invariance, and generates supercurrent. 
Actually, the Nambu's argument depends on the existence of the BCS-type particle-number mixed state, thus, if such a state is not physically allowed \cite{Peierls92,WWW1970}, the gauge invariance in the Meissner effect must be explained, differently. The new theory indicates that the BCS-type particle-number mixed state should be considered to be a mathematical tool to facilitate the calculation involving the electron pairing; the true superconducting state is actually given as a particle number fixed state. 

 In this section, we reexamine the gauge invariance problem in the BCS theory from the view point of the new theory.
 In the new theory, the gauge invariance in the Meissner effect is achieved by utilizing the gauge invariant gauge potential $(\varphi^{\rm eff}, {\bf A}^{\rm eff})$. 

First, we consider the gauge choice $\nabla \cdot {\bf A}^{\rm em}=0$ in Eq.~(\ref{BCSa}). In the new theory, the vector potential ${\bf A}^{\rm eff}$ appears in physical observables instead of ${\bf A}^{\rm em}$ and the choice of the gauge $\nabla \cdot {\bf A}^{\rm em}=0$ is compensated by the choice of $\nabla \chi$ in ${\bf A}^{\rm eff}$, thus, this condition can be used in the new theory as well.

Second, we take up the assumption, `${\bf A}^{\rm em}\!=\!0   \mbox{ if the magnetic field is zero}$', in Eq.~(\ref{BCSa}). This condition must be modified in the new theory since it is directly related to the observable current density.
The condition $\nabla \cdot {\bf A}^{\rm em}=0$ still leaves arbitrariness of the gauge for the zero magnetic field case. For example, 
\begin{eqnarray}
{\bf A}^{\rm em}={\bf A}_0={\rm const.}
\label{aconst}
\end{eqnarray}
 also fulfills the zero magnetic field and $\nabla \cdot {\bf A}^{\rm em}=0$. However, if this vector potential is employed, it yields the Meissner current for zero magnetic field.
 
 This problem is a very serious one in the calculation of the ${\bf q}=0$ Fourier component of the current density ${\bf j}$.
In the BCS theory, if ${\bf q} \rightarrow 0$ limit is taken, we have the following ${\bf q}={0}$ Fourier component of the current
\begin{eqnarray}
{\bf i}({ 0})=\Lambda {\bf a}^{\rm em}({ 0})
\label{i1}
\end{eqnarray}
where $\Lambda$ is a parameter, and ${\bf i}({0})$ and ${\bf a}^{\rm em}({0})$ are ${\bf q}={0}$ Fourier components of ${\bf j}$ and ${\bf A}^{\rm em}$,
respectively. This corresponds to Eq.~(5.26) in the BCS paper \cite{BCS1957}.
If we use a different gauge, this ${\bf a}^{\rm em}({0})$ can be removed. Thus, this current carrying state becomes a currentless state.

The problem here is related to the fact that the gauge degree-of-freedom may provide with a surplus whole system motion if the relation of the gauge of the gauge potential and the phase factor on the wave function are not intact as give in Eqs.~(\ref{gauge1}) and (\ref{gauge2}). If a surplus whole system motion exists, the conservation of the local charge may be violated. 
The removal of the surplus whole system motion is achieved in the process of obtaining $\nabla \chi$ in the new theory. 
On the other hand, the Ward-Takahashi relation is utilized in the standard theory.

Actually, if the condition in Eq.~(\ref{BCSa}) is replaced by
\begin{eqnarray}
\nabla \cdot {\bf A}^{\rm em}=0; \quad {\bf A}^{\rm eff}=0 \  \mbox{ if the magnetic field is zero.}
\label{BCSb}
\end{eqnarray}
the above-mentioned  problem is lifted. In this case, the constant vector potential is removed by adjusting $\chi$ as ${ \hbar \over {2q}}\nabla \chi=-{\bf A}^{\rm  em}=-{\bf A}_0$.
 
 A similar problem arises if we consider the situation where the magnetic flux quantization occurs. In this case, the vector potential in the magnetic field expelled region is given by
\begin{eqnarray}
{\bf A}^{\rm em}=-{\hbar \over {2e}}\nabla g
\label{zeroB}
\end{eqnarray}
where $g$ is an angular variable with period $2\pi$.
In this case, we have $\chi=-g$ from the condition ${\bf A}^{\rm eff}=0$; thus, zero current is obtained in the magnetic field expelled region with non-zero pure gauge.

\section{Concluding Remarks}
\label{section11}

When Schr\"{o}dinger solved the Schr\"{o}dinger equation for hydrogen atom, he required the wave function to be a single-valued function of the electron coordinate \cite{Schrodinger}.
 The single-valued requirement of the wave function is a postulate that can be rephrased as the existence of the basis $\{ |{\bf r} \rangle \}$ for the coordinate operator $\hat{\bf r}$ that satisfies 
 \begin{eqnarray}
 \hat{\bf r} |{\bf r} \rangle={\bf r}  |{\bf r} \rangle, 
 \label{singleO}
 \end{eqnarray}
 where $ {\bf r}$ is the eigenvalue uniquely determined by $ |{\bf r} \rangle$. 
 With this basis, the wave function for a state vector $|\varphi \rangle$ is given by $\langle {\bf r} | \varphi \rangle$, which must be single-valued with respect to the coordinate since $ {\bf r}$ is uniquely determined by $ |{\bf r} \rangle$ \cite{Koizumi2017b}.
 
Before the Schr\"{o}dinger equation was put forward by Schr\"{o}dinger, quantum mechanics was formulated as the Matrix mechanics by Heisenberg \cite{Heisenberg1925}. Schr\"{o}dinger showed that his version of
 quantum mechanics can be transformed into the Heisenberg's Matrix version by expressing the linear operators by matrices using the basis functions; then, the Schr\"{o}dinger's differential equation can be transformed into the matrix equation or the integral equation if the indices of the matrix elements are continuous \cite{Schrodinger2}.
 
However, von Neumann argued that these two forms are not equivalent; there are situations where differential equations cannot be simply transformed into integral equations, but require Dirac delta functions \cite{vonNeumann}. In this respect, the $\pi$-flux Dirac string is such an object. Actually, Dirac noticed the possible appearance of
 a phase factor in the displacement operator \cite{Dirac}, and also considered the possibile appearance of the singular phase factor in the wave function \cite{Monopole}. The Berry phase factor in the present work can be viewed as an example of such a phase factor.
 
Hohenberg and Kohn argued that the ground state can be obtained from the electron density alone \cite{Hohenberg1964}. However, their argument tacitly assumes the absence of singularities that might arise from many-body interactions and affect the phase of the wave function. When such singularities exist, we need to specify how to handle them. We assume that the basis satisfying Eq.~(\ref{singleO}) exists, and require that the wave function to be a single-valued function around the singularities.
 Then, the situation arises where the ground state cannot be obtained solely by the electron density alone, but requires the Berry connection.
 The present work indicates that one way to obtain it is to require the conservation of the local charge in addition to the single-valuedness of the wave function. Then, the so-called 'Bloch theorem' is violated, making it possible to generate supercurrent. 
  
The BCS theory uses the particle-number mixed state. There have been conflicting views on the use of such a state. Some researchers argue that it is unphysical thus should be considered as a mathematical tool to facilitate the inclusion of the electron pairing effects \cite{LeggettBook,Peierls92}; some consider that it is the essential ingredient of the theory to have the $U(1)$ gauge symmetry breaking \cite{Anderson}. In the present theory, the superconducting state is given as the particle-number fixed state in accordance with the former view. It is worth noting that the relation in Eq.~(\ref{Euler}) contains the subtraction of ``$1$'', which arises from the condition of the fixed total charge. This subtraction of ``1'' is also related to the topological structure of the real three dimensional space since the same relation holds as the Euler's theorem for a three dimensional object. This may mean that the local charge conservation is the condition to be imposed under the fixed total-charge constraint. If this is the case, requiring the conservation of the local charge using the particle number non-fixed formalism, which is employed in the $U(1)$ gauge symmetry breaking theory of superconductivity, is invalid.

In the new theory, the $\pi$-flux Dirac string is the necessary ingredient for the supercurrent generation. This can be considered as the $U(1)$ instanton, ${\bf A}^{\rm fic}=-{ \hbar \over {2e}} \nabla \chi$, ${\bf \varphi}^{\rm fic}={ \hbar \over {2e}} \partial_t \chi$, 
of Polyakov \cite{Polyakov1975}. In this respect, the superconductivity can be regarded as an instanton effect as in the chiral $U(1)$ gauge problem \cite{tHooft1976,Fujikawa2004}. In other words, the $U(1)$ gauge symmetry breaking in the standard theory is replaced by the appearance of the $U(1)$ instanton in the present theory.
  
 There is a connection between the Berry phase considered in the present work and the change of the $U(1)$ phase factor on the wave function when the gauge transformation is performed. This change is conveniently incorporated by using the effective gauge potential in materials $(\varphi^{\rm eff},{\bf A}^{\rm eff})$ since it is gauge invariant with respect to the choice of the gauge adopted in $( \varphi^{\rm em},{\bf A}^{\rm em})$ due to the fact that the arbitrariness in the gauge is absorbed in the Berry connection. 
 It is note worthy that an explanation is given to the long-standing puzzling problem of the `flux rule', the Faraday's induction formula is consist of one of the Maxwell equations and the Lorentz force calculation \cite{FeynmanII}, by using the effective gauge potential in materials \cite{FluxRule}. 
 
As far as the Rashba interaction is much smaller than the pairing energy and the phase variable is treated as a phenomenological parameter, the Ginzburg-Landau theory or the Bogoliubov-de Gennes equations will be used without modification.
However, the new origin requires the internal electric field of the  Rashba interaction for the occurrence of superconductivity. This may explain the fact that ideal metals like sodium does not show superconductivity since the screening of the electric field is efficient in such materials, suppressing the internal electric field too weak to occur superconductivity. 

It is also possible that the nontrivial Berry connection for many-body functions may arise from other degree-of-freedom than spin; for example,
orbital degree-of-freedom may give rise to it. In this respect, it is noteworthy that the band crossings or Lifshitz transitions are argued to be relevant to the superconductivity in the pressurized sulfur hydride \cite{Jarlborg2015,Jarlborg2016}.
 
\begin{acknowledgements}
 Part of the present work was conducted during the author's sabbatical stay at Lorentz Institute for theoretical physics, Leiden University, the Netherlands. 
He thanks the members of the institute for their hospitality.
\end{acknowledgements}


\bibliographystyle{spphys}       
\end{document}